\documentclass[journal]{vgtc}
\ifpdf
  \pdfoutput=1\relax                   
  \usepackage{graphicx}                
  \DeclareGraphicsExtensions{.pdf,.png,.jpg,.jpeg} 
\else
  \usepackage{graphicx}                
  \DeclareGraphicsExtensions{.eps}     
\fi%
\graphicspath{{./figures/}{./pictures/}{./images/}{./}} 

\vgtcinsertpkg

\usepackage{tcolorbox}
\usepackage{microtype}                 
\PassOptionsToPackage{warn}{textcomp}  
\usepackage{textcomp}                  
\usepackage{mathptmx}                  
\usepackage{times}                     
\usepackage{cite}                      
\usepackage{tabu}                      
\usepackage{booktabs}                  
\usepackage{wrapfig}
\usepackage{calc}
\usepackage{todonotes}
\usepackage{comment}
\usepackage{multirow} 
\usepackage{caption}
\usepackage{subcaption}
\usepackage{epstopdf}
\usepackage{amsmath}
\usepackage{array}
\usepackage{tikz}
\usepackage{xspace}
\usepackage[percent]{overpic}
\usepackage{enumitem}
\usepackage{xcolor}

\usepackage[export]{adjustbox}

\usepackage{fancybox, graphicx}

\newlength\myheight
\newlength\mydepth
\settototalheight\myheight{Xygp}
\newcommand*\wordimg[1]{%
  \settototalheight\myheight{Xygp}%
  \settodepth\mydepth{Xygp}%
  \raisebox{-\mydepth}{\includegraphics[height=\myheight]{#1}}%
}

\newcommand{\changes}[1]{\protect\textcolor{Blue}{#1}}
\renewcommand{\changes}[1]{#1}

\newcommand{\tochange}[1]{\protect\textcolor{black}{#1}}

\definecolor{VividBurgundy}{RGB}{159,29,53}
\definecolor{mscolor}{rgb}{0,0,0.7}
\definecolor{picolor}{HTML}{2a9d8f}
\definecolor{jccolor}{RGB}{159,89,53}
\definecolor{tmcolor}{rgb}{0.7,0,0}

\newcommand{\codingdecisions}[1]{\textit{#1}}
\newcommand{\codingrules}[1]{\textcolor{black}{#1}}

\newcommand\msc[1]{{\color{black}#1}}
\newcommand\drpi[1]{{\color{black}#1}}
\newcommand\jc[1]{{\color{jccolor}[JC: #1]}}
\newcommand\jcjc[1]{{\color{jccolor}#1}}
\newcommand\tomo[1]{{\color{tmcolor}TM: #1}}
\newcommand\tic[1]{\textcolor{BrickRed}{[TI: #1]}}

\newcolumntype{L}[1]{>{\raggedright\arraybackslash}p{#1}}
\newcolumntype{C}[1]{>{\centering\arraybackslash}p{#1}}
\newcolumntype{R}[1]{>{\raggedleft\arraybackslash}p{#1}}

\newcommand{\eg}{e.\,g.}
\newcommand{\ie}{i.\,e.}

\newcommand{\vislength}{``generalized bar representations''\xspace}
\newcommand{\vispoint}{``point-based representations\xspace''}
\newcommand{\visline}{``line-based representations''\xspace}
\newcommand{\visnodelink}{``Node-link trees\discretionary{/}{}{/}graphs, networks, meshes''\xspace}
\newcommand{\visarea}{``generalized area representations''\xspace}
\newcommand{\vissurface}{``sur\-face-based representations and volumes''\xspace}
\newcommand{\visgrid}{``generalized matrix and grid''\xspace}

\newcommand{\visglyph}{``glyph-based representations''\xspace}
\newcommand{\visschematic}{``sche\-ma\-tic representations and concept illustrations''\xspace}

\newcommand{\viscolor}{``continuous color-based representations''\xspace}

\newlength{\pictureheight}

\newcommand{\mytitle}{%
Not As Easy As You Think---Experiences and Lessons Learnt\\ from Creating a Visualization Image Typology%
}

\graphicspath{{IEEEtran/figures/}{IEEEtran/figuresExamples/}{./}{IEEEtran/FigureVCExamples/}} 

\newcommand{\imagesfigSeven}{6,833\xspace}
\newcommand{\totaltypelabels}{9,039\xspace}
\newcommand{\totalhardnesslabels}{13,666\xspace}
\newcommand{\totaldimlabels}{7,181\xspace}

\newcommand{\papersSeven}{695\xspace}

\newcommand{\numguischematicCodeSingletons}{2,619\xspace}

\newcommand{\lineprop}{$28.1\%$}

\newcommand{\patternprop}{$2.9\%$}

\newcommand{\barcon}{$67\%$\xspace}
\newcommand{\pointcon}{$57\%$\xspace}

\newcommand{\glyphcon}{$35\%$\xspace}

\newcommand{\textcon}{$39\%$\xspace}

\newcommand{\guicon}{$70\%$\xspace}

\newcommand{\imagesfunction}{4,214\xspace} 

\newcommand{\totalfunctiontypelabels}{6,299\xspace}

\newcommand{\numschematicOrg}{1,919\xspace} 
\newcommand{\numguiOrg}{825\xspace}  

\newcommand{\inlinevis}[3]{\raisebox{#1}[0pt][0pt]{\includegraphics[height=#2]{#3}}}

\onlineid{0}
\vgtccategory{Research}
\vgtcpapertype{please specify}

\title{\mytitle}
\author{Jian~Chen, Petra~Isenberg, Robert~S.~Laramee, Tobias~Isenberg, Michael~Sedlmair, Torsten~M{\"o}ller, Han-Wei~Shen}
\authorfooter{
\item J. Chen and H.-W. Shen are with The Ohio State University, USA.
E-mail: \{chen.8028\,$|$\,shen.94\}@osu.edu.
\item P. Isenberg and T. Isenberg are with Université Paris-Saclay, CNRS, Inria, LISN, France. E-mail: \{petra.isenberg\,$|$\,tobias.isenberg\}@inria.fr.
\item R. S.\ Laramee is with the University of Nottingham, UK.
E-mail: robert.laramee@nottingham.ac.uk.
\item M. Sedlmair is with University of Stuttgart, Germany.
E-mail: michael.sedlmair@visus.uni-stuttgart.de.
\item T. M{\"o}ller is with University of Vienna, Austria.
E-mail: torsten.moeller@univie.ac.at.%
}

\manuscriptnote{\vspace{-2\baselineskip}\vspace{-4pt}}

\shortauthortitle{Biv \MakeLowercase{\textit{et al.}}: \mytitle}

\abstract{We present and discuss the results of a two-year qualitative analysis of images published in IEEE Visualization (VIS) papers. 
Specifically, we derive a typology of 13 visualization image types, coded to distinguish visual designs and several image characteristics. 
The categorization process required much more time and was more difficult than we anticipated.
The resulting typology and image analysis may serve a number of purposes: 
to study the evolution of the community and its research output over time, 
to facilitate the categorization of visualization images for the purpose of research or teaching, 
to identify visual design styles, or 
to enable progress towards standardization in visualization.
In addition to the typology and image characterization, we provide a dataset of \imagesfigSeven~tagged images and an online tool 
that can be used to explore and analyze the large set of tagged images.  
The tool and data set enable a close examination of the diverse visualizations used and how they are published and communicated in our 
community.
} 

\keywords{Visualization, classification, images, typology}

\CCScatlist{ 
 \CCScat{K.6.1}{Management of Computing and Information Systems}%
{Project and People Management}{Life Cycle};
 \CCScat{K.7.m}{The Computing Profession}{Miscellaneous}{Ethics}
}

\teaser{
  \centering
  \includegraphics[width=\linewidth]{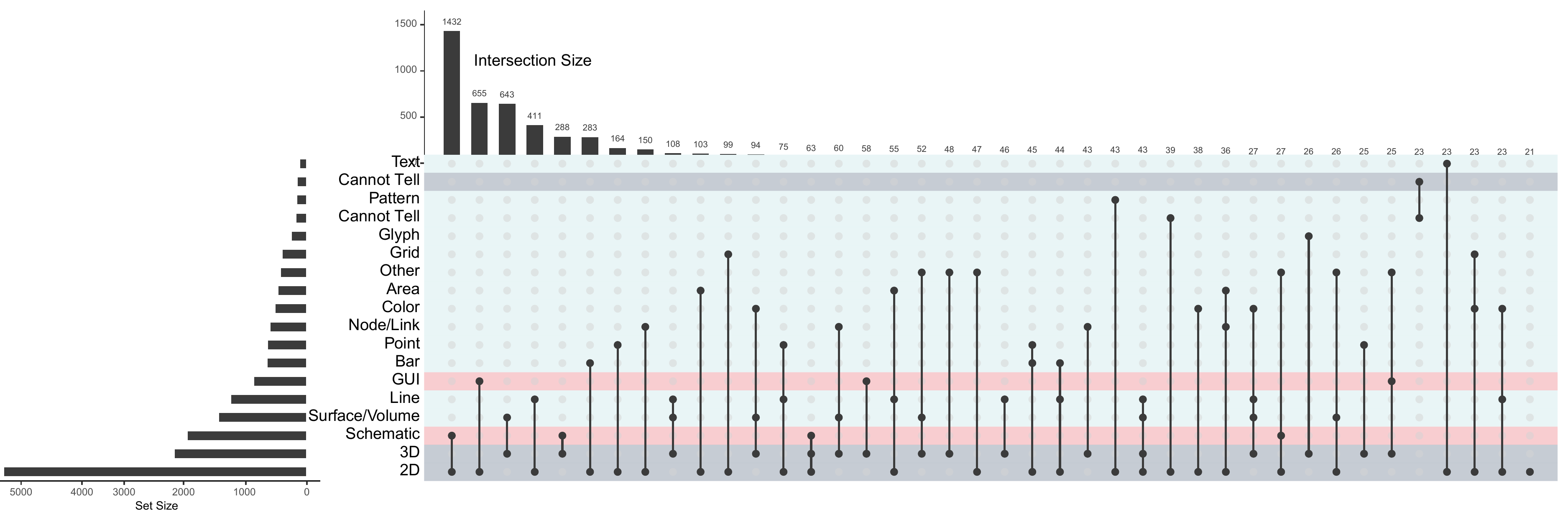}
  \caption{Coding results from categorizing IEEE VIS paper images according to: visualization types (light blue), their dimensionality (dark blue), and additional image categories (red). 2D Schematics are the most common type of figure in IEEE VIS publications, followed by 3D surface/volume renderings.%
  }
  \label{fig:teaser}
}

\begin{document}

\firstsection{Introduction}
\label{sec:introduction}
\maketitle

The visualization research space has been studied from a variety of angles.  Some started by considering specific data types (TimeVis~\cite{aigner2011visualization}, TreeVis~\cite{pandey2021state}),  others looked at keywords~\cite{isenberg2016visualization}, evaluations~\cite{isenberg2013systematic,lam2011empirical}, 
topic modeling~\cite{jiang2016text}, interaction~\cite{yi2007toward}, or taxonomies of tasks and activities~\cite{dimara2021critical,amar2005low}. Describing and classifying the challenges, artefacts, research methods, and theories in a research field is a difficult endeavor but this work is nevertheless very important for a variety of tasks. For example, a classification of visualization techniques can be extremely helpful when planning any research activity that requires a systematic coverage of the space of existing visualization. For example, when developing a new research method applicable to a variety of visualization techniques, it is important to test the method and ground the design on a broad set of techniques or variations of a specific technique~\cite{Sedlmair:2012:DSM}. Similarly, when writing an overview article, a textbook, or lecture series about visualization---classifications can illuminate coverage of techniques, to show the variety of approaches to a specific techniques, or to identify aspects that need further attention. When trying to understand the historical evolution of the visualization field, it may be similarly useful to systematically consider the types of artefacts and research produced through the lens of a classification. In all three cases more broadly, a systematic organization can help to ensure coverage of a research space, identify outliers, structure discussions, and potentially even open up venues for future work \cite{lohse1994classification, purchase2018classification}. 

While many characterizations of visualization tasks, visualizations, keywords, or topics exist, what has not been systematically attempted so far is a bottom-up approach that starts with the visual artefacts published, communicated, and discussed in the community. We provide such a view on our research space and publication practices by systematically analyzing images published throughout the entire 30 year history of the IEEE Visualization conference---as the longest running venue for the publication of novel representation types, the evaluation of existing techniques, or the development of visualization systems (among other types of contributions). Specifically, we coded \imagesfigSeven~figures from \papersSeven~papers published in IEEE VIS (VisWeek) 1990, 1995, 2000, 2005, 2010, 2015, and 2020, a subset of the VIS30K dataset~\cite{chen2021vis30k}. Our initial goal was to study how visualizations were used to communicate research in the community. Throughout our two-year collaboration and lively discussions, this goal evolved toward establishing a broader typology of the visualizations we saw as well as a description of how these visualizations are used in visualization research publications. Our discussions focused on the visual appearance of these visualizations without consulting the captions and reading the authors' intents in the papers. We discussed whether the images showcased a certain type of visual design, a system or GUI element, or were schematics meant to explain workflows or processes.  Our final code set describes 13 image types and their perceived dimensions (\autoref{fig:teaser}).
We find that the largest categories were schematic representations, surface-based techniques \& volumes, line-based-technique, and GUIs. Together, these top four categories account for
\tochange{$73\%$ (4,986 out of 6,833)}  
of the coded images.

By relying on our experience in visualization research, teaching, and practice we initially assumed the coding process would be relatively straightforward. 
Having studied visual encoding principles as put forward by Bertin~\cite{bertin1983semiology}, 
MacKinlay~\cite{mackinlay1986automating}, and others, the design space of visualizations, in theory, seemed more or less clearly defined.
However, it quickly became clear that visual designs ``in the wild'' (even with our constraint to the academic visualization community) display a great variety and demonstrate tremendous creativity, to the extent that the complexity of coding these charts is a major challenge of our work.
There are, for instance, no standard definitions of many visual designs, \eg, glyphs and grid-based techniques, that are specific enough to foster a clear-cut coding process of visualization images. Given our shared background in visualization, we were also surprised about the large role that individual differences and interpretations played. And for some images, even after intense discussions, their categorization remained ambiguous and uncertain. %
%
In summary, we contribute:
\begin{itemize}[topsep=1pt,itemsep=1pt,partopsep=0pt,parsep=0pt]
    \item a novel typology of visualization images consisting of 13 categories,
    \item  the coding dataset and quantitative analysis of \imagesfigSeven~IEEE VIS (VisWeek) images based on the typology,
    \item \msc{a discussion of our process, failed attempts, and 
    coding ambiguities
    in deriving the typology, and}
    \item an open web-based tool to explore the image dataset.
\end{itemize}




\section{Related Work}
\label{sec:relatedWork}
Past work on visualization categorizations is related to our own, as well as work that analyzes research figures. We review these areas next.


\subsection{\changes{Categorization as an Analogy of ``What is it \textit{like}?'' Association: A Brief History}}

\changes{
Categorization represents any grouping based on similarity~\cite{bar2007proactive}, originated from Plato in philosophy~\cite{givon1986prototypes}. Wittengenstein~\cite{wittgenstein2010philosophical} argues that people are fluid about categories: different people may give different answers and the same person may give us at different time different answers. As a result, a top-down categorization that draws clear boundaries of shared properties to those of other categories, is too strict to represent how people understand categories.
Rosch~\cite{rosch1973natural} later updated the notation of categorization using a data-driven solution where categories become \textit{prototypes} from bottom-up clustering of similar instances. These prototypes can get clustered again to form a hierarchical categorization. Psychologists (\eg, Medin and Schaffer~\cite{medin1978context}, Nosofsky~\cite{nosofsky1986attention}, and Krushki~\cite{kruschke2020alcove}) have gone further
and argue for what they call the \textit{example-based theory} of categorization, where humans store 
\textit{instances} and 
\textit{examples} of these; similarities and differences among which let us see and learn \textit{associations} between these examples. For things closer together, 
they \textit{look like} a clustering of learned examples and belong to the same category. 
Human observations of similarity depend on high-order structure~\cite{wang2004image} (thus
cannot be represented by low-level features~\cite{zhang2018unreasonable}), and are context-dependent~\cite{medin1993respects} (thus categorical boundaries can thus be fuzzy). In this work we also stepped away from the top-down categorization paradigm and tried to categorize visualizations as a bottom-up association of what we see from these images.}

\subsection{Visualization Categorizations}
Images have been categorized \changes{based on how they are constructed, rather than how they are seen by viewers.} Textbooks, in par\-ti\-cu\-lar, often rely on categorizations to structure their content \cite{rees:survey}. While early books such as Brinton's \cite{Brinton:1939:GP} were collections of graphical representations in use, modern textbooks regularly use one of a few structures:

\textbf{Focus on construction rules \changes{for design purposes}.} 
A first and highly influential approach to characterizing visual designs was Bertin's visual semiotics~\cite{bertin1983semiology}. 
He discussed the fundamental building blocks of images that are modified by visual variables (visual channels), which encode data. 
Similar in spirit, several proposals have been made to describe visual designs through the lens of a visual language with a set of syntactic rules. Examples include Wilkinson's Grammar of Graphics \cite{Wilkinson:2005:GrammarOfGraphics}, Engelhardt's Language of Graphics \cite{Engelhardt:2002:LanguageOfGraphics}, or Mackinlay's automatic design \cite{mackinlay1986automating}.  
Applying rules formulated in a visual language can yield a broad range of
visualization designs \cite{munzner2014visualization} and several visualization tools and libraries are based on them, \eg, Tableau \cite{mackinlay2007show}, D3 \cite{bostock2011d3}, or Ve\-ga-lite \cite{satyanarayan2016vega}. Others, such as Tufte's Envisioning Information \cite{Tufte:1990:Envisioning} differentiate techniques by higher-level construction rules such as small multiples, or principles such as layering and separation. 

What unites these approaches is that they attempt to describe \emph{how} to construct a visualization but do not focus on \emph{what} the end-product of the construction looks like.  Different sets of rules may lead to images that \emph{look} very similar.
\changes{Intriguingly, these construction rules cannot name a category nor tell apart visualization categories. For example, we cannot use length, area, and orientation to differentiate a bar chart from a pie chart. In another words, humans cannot always rely on abstract definitions or shared drawing entities to predict categories~\cite{wittgenstein2010philosophical, rosch1973natural}.}
We used an inverse approach in that we took existing images and attempted to describe their visual appearance. By using this approach, our characterization incorporates our assessment of what is important in a visualization; this assessment is certainly related to encoding of marks and channels without necessarily considering what data is encoded.
\changes{This approach enables us to generate high-level categories beyond marks and channels.}

\textbf{Focus on data \changes{types}.}
Many researchers have categorized visualization techniques based on the type of data they show. This approach makes sense as, in a typical iterative visualization design process, data are systematically mapped, winnowed, and refined to visual encoding \cite{Sedlmair:2012:DSM}.
Ward et al.~\cite{Ward:2015:IDV}, \eg, categorize visualization techniques for spatial data, geospatial data, time-oriented data, multivariate data, trees, graphs, and networks, and text and document visualizations. Heer et al.'s visualization zoo \cite{Heer:2010:TVZ} classifies time series, statistical data, maps, hierarchies, and networks.  
Brodlie \cite[pp. 40ff]{Brodlie2012scientific_chap3} classifies techniques into those for point, scalar, vector, and tensor data.  Similar to the first approach, these categorizations focus on how to construct a visualization and do not focus on how to describe the visual appearance of an image. Compared to these characterizations where data is input and visual images are the output, we attempt to characterize visualizations without necessarily knowing the characteristics of the data that led to the final image. 
For example, we make no distinction between a line chart that shows temporal data and one that shows, \eg, a physical measurement such as voltage that was sampled in some arbitrary sequence but plotted in a meaningful way from low to high values. 

\textbf{Focus on task and analysis question \changes{types}.}
Another set of textbooks introduces visualizations by linking representation and analysis tasks/questions. Fisher and Meyer~\cite{Fisher:2018:MDV}, for example, group techniques such as histograms and boxplots under the analysis question of ``showing how data is distributed.'' Maciejewski~\cite{Maciejewski:2011:DRT} also takes this approach in his grouping of techniques.
Again, a focus on analysis questions considers a-priori criteria to choose and categorize visualization techniques in the same vein as data and construction rules do. As a consequence, visually similar techniques are considered in separate categories; for example, Fisher and Meyer~\cite{Fisher:2018:MDV} categorize bar charts under ``visualizations that show how groups differ'' and histograms under ``visualizations that show how data is distributed.'' Again, our approach attempts to uniquely identify images from the standpoint of having been created already, without necessarily knowing what data they show, how they have been constructed, or what tasks they were meant to serve. This allows us to group visually similar techniques together and only later to consider other aspects in which they differ.

\changes{\textbf{Most categories are functional.} We 
are certainly not the first to realize the feature differences between \textit{what we design} and \textit{what we see}. The computer vision community has realized that many descriptors (\eg, Canny edge detector~\cite{canny1986computational}, orientation map~\cite{malik1997computing}, and HOG algorithms~\cite{felzenszwalb2010object}) use features that do not align with what humans see, and therefore cause some computer vision algorithms to fail. Recent artificial intelligence algorithms are better able to assign categories to items because the categories are treated as a continuous space of related high-level concepts
\cite{chen2019looks, malisiewicz2008recognition}. 
In the space of artificial intelligence categorization, our results are thus useful for future grouping and classifying new visualization techniques (akin to ImageNet in computer vision~\cite{krizhevsky2017imagenet}). 
Nonetheless, we also show the diversity of the data: even under a single category.
Our collection combines various variable appearances, spatial arrangements, appearances (color), compositions, and viewpoints. 
}



\subsection{The Role of Graphs in Scientific Communication} 

Though our perspectives on how to categorize visualizations is different from that of many textbooks, 
our method of studying figures in evaluating scientific advances has been used before. 
Latour~\cite{latour2012visualisation} laid out graph features that make them essentially a pervasive form of visualization and a specialized vocabulary for transforming and analyzing data to represent scientific findings. The pervasiveness and centrality of scientific figures led Latour to conclude that scientists exhibit a ``graphical obsession'' and indeed to suggest that the use of graphs distinguishes scientific domains. 
In the visualization domain, our work is most closely related to Borkin et al.'s~\cite{borkin2013makes} work towards studying what makes images memorable. In that work, the authors suggested a taxonomy of techniques that is a mix of encoding (area, bar, \dots), data (network, tree, \dots), and analysis-focused (distribution) categories. The authors asked students to annotate 2,070 single-panel visualizations using their taxonomy and derived an annotated set of images. Our work differs, however, in several aspects from Borkin et al.'s: we describe and discuss the process of deriving our categorization and the inherent difficulties, we included images with multiple visual encodings, we focus on 
visualization articles as a source and do not study memorability as a final goal of our work.
Our approach is one of the few that focuses on images only.
Visuals are one of the most 
essential outputs from the visualization community (as opposed to data or tasks). 
We consider them to be very important since they are at the center of our work.
Our project thus facilitates the classification, exploration, and analysis of our own fundamental content through a new image-centric lens.

\section{The Image Coding Process}
\label{sec:process}

%

\begin{table*}[!tb]

  \caption{The main visualization type, function, and dimensionality codes used in our review. Additional codes not listed here were ``I cannot tell'' to label images that had unclear techniques or dimensionalities.}
  \label{tab:12schema}
  \scriptsize%
	\centering%
	\renewcommand{\arraystretch}{1.4}
  \begin{tabu}{@{}c@{}R{0.175\textwidth}L{.455\textwidth}L{.295\textwidth}@{}}
  \toprule
  &\textbf{Visualization Type Codes} &  \textit{Description} & \textit{Examples}\\
  \midrule
(1) &Generalized Bar Representations & 
Graphs that represent data with rectangular bars whose heights or lengths are proportional to the values they represent. 
& bar charts, stacked bar charts, histograms, box plots, sunburst diagrams.
 \\
 (2) &Point-based Representations & Representations that use point marks to represent data samples & scatterplots (2D/3D), point clouds, dot plots, bubble charts.\\
 (3)& Line-based Representations & Representations in which lines, edges, or curves represent data samples. Lines can depict surface features or data values.
 & line charts, parallel coordinates, contour lines, radar/spider charts, streamlines, or tensor field lines.\\
(4)& Node-link Trees/Graphs, Networks, Meshes & 
 Representations using points for nodes/points and explicit connections to convey relationships between data values & node-link diagrams, node-link trees, node-link graphs, meshes, arc diagrams, sankey diagrams.\\
(5)& Generalized Area Representations & Representations with a focus on areas of 2D space or 2D surfaces including sub-sets of these surfaces. Areas can be geographical regions or polygons whose size or shape represents abstract data. 
 & (stacked) area chart, 
    area chart,
    streamgraph,
    ThemeRiver,
    violin plot, 
    cartograms,
    rideline chart,
    voronoi diagram,
    treemaps,
    pie chart.\\
 (6) & Surface-based Representations and Volumes & 
  Representations of the inner and/or outer features and/or boundaries of a continuous spatial phenomenon or object in
  3D physical space or 4D space-time, or slices thereof.  
 & terrains, 
 isosurfaces, 
 stream surfaces,
 volume rendering using transfer functions, 
 slices through a volume (\eg, X-ray, CT slice).\\
 (7)& Generalized Matrix / Grid & 
 Representations that position data in a \textit{discrete} grid structure.  
 The grid can vary in resolution, is typically rectilinear but can use other shapes such as hexagonal grids etc.,
  &
  network matrices, 
  discrete density maps, 
  scarf or strip plots.\\
  \\
 (8) & Continuous Pattern-based Representations & 
  Representations of continuous data along planes and surfaces, typically for vector and tensor fields. Representations frequently use texture-mapped imagery to describe variations in orientations, directions, or flow. 
  & Line Integral Convolution (LIC), Spot Noise, Image-Space Advection (ISA), Image-Based Flow Visualization (IBFV)  \\
 (9) & Continuous color-based Representations & Representations that use a primary  encoding where the hue or brightness or saturation encodes quantitative values on a continuous surface. Color-mapping is systematic, thus is not as a result of illumination or an author-chosen categorical representation. 
  & 
  pixel heatmaps, color-mapped surfaces.\\
 (10) & Glyph-based Representations & 
    Multiple small independent visual representations that depict multiple attributes (dimensions) of a data record. Placement is usually meaningful and typically multiple glyphs are displayed for comparison.  
   & 
   Star glyphs, 3D glyphs, Chernoff faces, vector field glyphs\\
 (11) & Text-based Representations &  Representations of data (usually text itself) that use varying properties of letters and words such as font size, color, width, style, type to encode data.
  & 
 tag clouds,
 word trees,
 parallel tag clouds,
 typomaps.\\
 \midrule
 & \textbf{Function Codes} &  \textit{Description} & \textit{Examples} \\
 \midrule
 A. & GUI Screenshots or GUI Photos & 
  Images that show a system or user interface. 
  & a photograph of a person sitting in front of a given system, 
  a figure containing GUI features such as windows, icons, cursor, and pointers (WIMP), or non-WIMP VR/AR interfaces.\\
 B. & Schematic Representations, Concept Illustrations & 
 Often simplified representations showing the appearance, structure, or logic of a process or concept. 
  & workflow diagrams, algorithm diagrams, sketches.\\
 \midrule
 &\textbf{Dimensionality Codes} &  \textit{Description} & \textit{Examples} \\
 \midrule
 & 2D & Flat representations, no specific depth codes added to renditions. & Most statistical charts, most maps, \dots
 \\
 & 3D & Representation with specific depth cues that achieve the perception of 3D (shading, perspective, lighting, …). & Most volume renderings, \dots \\

  \bottomrule
  \end{tabu}\vspace{-1ex}%
\end{table*}



\changes{Our image coding is an ambitious project. 
We list a set of high-level goals we strive for and the process to reach these goals.}
\tochange{We begin our discussion with a temporal overview of our process to classify images and our systematic methodology. Our process became one of open and axial coding \cite{charmaz2006constructing} while we continuously updated, drew connections between, and refined our codes as we analyzed more data.}
\subsection{\changes{Goals of our Image Coding}}

\changes{To summarize our previous discussion, our categorization of images according to ``what we see'' has four concrete goals:}

\changes{
\textbf{Provide an alternative viewpoint:} Rather than categorizing visual designs from underlying data, construction rules, or functions we provide a categorization based on the visual content of images alone. This approach offers a new viewpoint that can serve to compare and complement other categorizations and puts the focus on the diversity of aesthetics and other visual properties within a single category. }

\changes{\textbf{Collect experiences concerning the difficulties of categorizing visualization images:} We document our  multi-stage process to derive a relatively high-level categorization of images and describe inherent uncertainty, failed attempts, and current limitations. 
We also describe how difficult it can be to understand images that have been taken out of the context of the text and captions. }

\changes{\textbf{Provide a small set of broad categories:} We purposefully wanted to create a classification with only a few categories that would remain manageable given the detailed and often complex types of images produced in the Visualization community. These categories needed to capture the diversity of design approaches, rendering methods, algorithms, or view point selections within a category. }

\changes{\textbf{Provide data and explore the use of images in the visualization community.} This exploration can give valuable insight in the changing practices of communication and research in our community.}



\subsection{Visualization Image Data Source}



\noindent We developed our classification 
using
the VIS30K \cite{chen2021vis30k} collection of images and its associated VisPubData \cite{isenberg2016vispubdata} meta data.
This dataset largely represents visualization as a field because it contains every visualization image published at 
IEEE VIS 
(including Visual Analytics, Information Visualization, and Scientific Visualization)
since 1990.
Based on our pilot studies (described below) it became clear that we would not be able to classify all 30,000 images.
Thus we chose to code images in five-year intervals for our primary study,  starting with 1990 and up until 2020 (inclusive).
This led to a set of \imagesfigSeven
figures 
from \papersSeven IEEE VisWeek/VIS full papers (including case studies), which we analyzed. 
In each phase of the work, seven experienced coders, all co-authors of this manuscript, classified subsets of the images.

\subsection{Image Classification Process}
Included in our process description are the approximate start dates and duration of each phase.

\textbf{Phase 1---Initial image classification based on keywords}
\tochange{(circa Mar.\ 2020 start, approx.\ 1 month)}:
We began our work with a focus on \emph{visualization techniques} where we considered that technique names such as treemaps, parallel coordinates, etc.\ could well describe the content of the images we were analyzing. To improve objectivity and reduce bias, we wanted to tag images with the most common technique names used in the community as extracted from author keywords used for VIS papers. We ranked the author keywords extracted in prior work \cite{isenberg2016visualization} to derive
the initial top-21 keywords for specific techniques.
In addition to the encoding techniques used in each image, we added two code categories that seemed important to be able to describe visualizations and communication practices in the community: the rendering dimensionality (\ie, 2D or 3D) and the functional purposes of creating the image
(\ie, the reason why the authors created each image, for example: illustration of a visualization technique, experiment results, or screenshot of GUI.) See \autoref{sm:process} for the initial keyword list and functions.
\drpi{The initial code set, thus, included 28 codes}.

\textbf{Phase 2---Initial pilot coding}  
\tochange{(circa Apr.\ 2020 start, approx.\ 2 months)}: 
To test our initial code set, each coder categorized visualization images from the year 2006. 
Visualization images from the year 2006 were used in our pilot study only.
We subsequently introduced new technique codes by merging techniques that had been tagged ``other,'' 
and added ``schematic diagram'' to the list of image purposes, which resulted in 22 technique codes.  
We discussed the definition of these terms and gave all coders written instructions and example images from each category.
Each figure was labeled initially by one coder in this stage and validated by a second coder. We based the validation assignment of the second coder on their respective expertise, to verify all images included and excluded in a specific category 
(\eg, volume rendering was coded and verified by experts with a sustained track record in volume graphics).
With these steps we removed false positives, avoided false negatives, and ensured image classification consistency. 

\textbf{Phase 3---Consolidation: Seeing by association and analogies} 
\tochange{(circa Jun.\ 2020 start, approx.\ 5 months)}:
In Phase 3, we discussed what worked well and what did not, and why. 
\drpi{The codes focused on visualization techniques quickly became difficult to apply as the number of techniques grew increasingly large. 
We had difficulty defining when a technique should receive its own code or be covered under ``other.'' Also, some technique names were different but pointed to visually similar designs. 
Point clouds and 3D scatterplots, \eg, both render points according to underlying coordinates in 3D space, with the main exception that scatterplots typically include reference structures such as axes and gridlines.}
%
These conflicts in turn lead us to re-frame our code set using higher-level (more general) visualization type codes. 

We
decided to focus on describing the main perceptual (or visual) characteristic of a given visualization and to create codes that enable the viewer to distinguish graphical similarities and differences. 
We thus re-grouped and merged the codes sharing similar visual characteristics into a more general code. We put \emph{isosurface}, \eg, into a more general \emph{surface-based techniques} category and grouped \emph{point clouds} and \emph{scatterplots} into a more abstract \emph{point-based techniques} category.

This consolidation process resulted in 12 high-level visualization type codes that ultimately became part of our final set shown in \autoref{tab:12schema}. \drpi{We reduced our image function category into just two codes: GUI (Screenshots) and Schematics as these were visually identifiable without requiring knowledge of the underlying data.} Both categories represent the codes from other encoding categories from Phase~1, which did not appear in our technique-focused codes. 
Given the challenges to code the visualization images we had encountered, we also
decided to collect subjective ratings of difficulty (easy, neutral, hard).


\textbf{Phase 4---2\textsuperscript{nd} Pilot Visualization Typology} 
(circa Dec.\ 2020 start, approx.\ 3 months): 
In the first two months after having arrived at the new code set, we discussed, debated, and coded two sets of 50 (\ie, 100 in total) randomly chosen images from our seven-year target image data set to calibrate our collective understanding.
During this exercise we clarified code definitions and discussed ambiguities. 
We assigned these 100 images to all of us for quality control as well to align our decisions and discuss potential pitfalls with our new code set.
We used a dedicated web-based coding tool (\autoref{sm:web-interfaces}) for the coding in this phase. 
At the end of this phase, our project was already one year old. We had almost weekly meetings and discussions throughout this time and
were now confident that we had largely reached consensus on the general typology shown as the top 13 codes (except color) of \autoref{tab:12schema}.

\textbf{Phase 5---Result Coding and Validation} 
(circa Apr.\ 2021 start, approx.\ 6 months): 
In this phase we coded all \imagesfigSeven~images in our chosen dataset based on the refined definitions and characterizations, using largely the same coding tool as before (\autoref{sm:web-interfaces}). \drpi{We used the 12 visualization type codes, 2 function codes (GUI, Schematics), 2 codes for the dimensionality (2D vs.\ 3D), and 3 difficulty codes (easy, neutral, hard) that meant to capture how hard the categorization had been for the coder.
We first looked at the function of an image. If the image showed either a schematic or a GUI we assigned the respective code. If not, the function was implicitly considered to be a ``visualization example'' and we proceeded to assigning a visualization type code. Both categories shared an ``I cannot tell'' code that was assigned when neither one of the two explicit functions nor a visualization type could be assigned. In addition, coders could freely add new codes for visualization or image types when they found something new that was not covered by the existing codes. For the dimensionality we also added a code called ``I cannot tell'' which could be checked when coders were unsure whether the visualization was a 2D or 3D rendering. Coders could assign multiple types and dimensionalities to one image which was necessary because many images showed more than one visual design. }

In the coding process, we also recoded the 100 images we had previously pilot-coded during Phase 4. To ensure high-quality results and more clearly capture potential difficulties in applying the codes two coders were randomly assigned to each image. This phase was lengthy and laborious due the number of images we classified. \drpi{We met regularly to resolve further questions that arose during coding and to augment or clarify code descriptions further.}

\textbf{Phase 6---Verification} (circa Oct.\ 2021 start, approx.\ 6 months): 
In the final phase, the two coders assigned to each image worked to reach an agreement when their applied codes did not match. 
For this purpose we developed two more web-based visual interfaces that focused on conflict resolution and giving and overview of applied codes (\autoref{sm:web-interfaces}).
We then filtered the results such that only the inconsistently coded images were shown, and we used this process to resolve all disagreements.
This verification was also a lengthy and laborious process requiring consistent discussions. 
We discussed difficult and ambiguous cases as a team every week until we could agree on a solution (we describe some of the most difficult classifications in more detail in \autoref{sec:ambiguous}). 
As part of this discussion we added a 13\textsuperscript{th} visualization type, \emph{continuous color-based encodings}, listed as point (9) in \autoref{tab:12schema}. 
We also refined our definitions further. 
Consequently, we went through all previously coded images again to check if they had to be re-coded for consistency, and resolved potential resulting disagreements as part of our discussion process.


\section{Results}
\label{sec:13vcs}




\begin{figure}[!bt]
    \centering
    \includegraphics[clip, trim={0 0 12 20}, height=0.49\columnwidth]{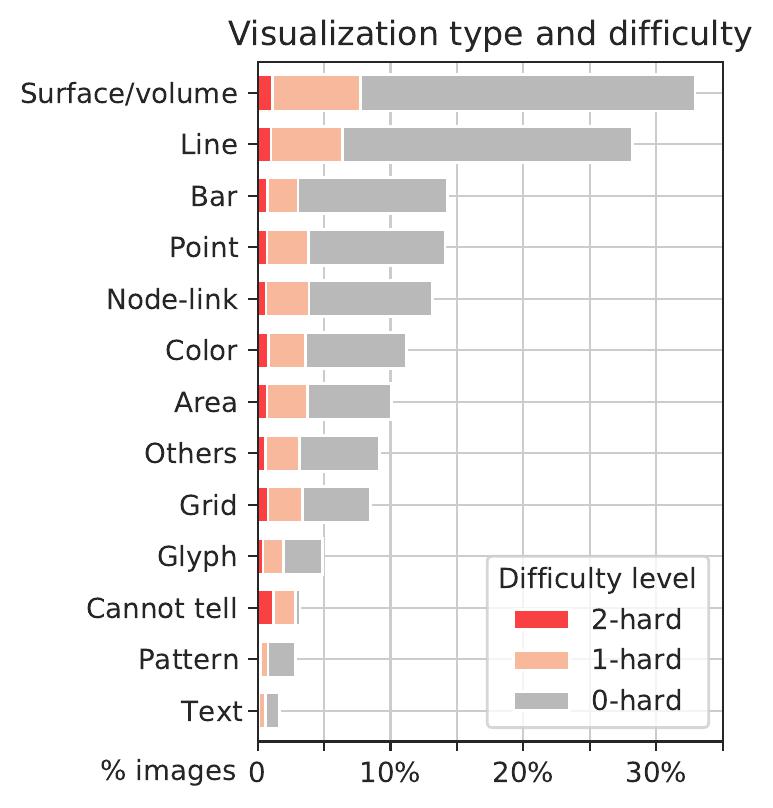}
    \hfill
    \includegraphics[clip, trim={0 0 12 20}, height=0.49\columnwidth]{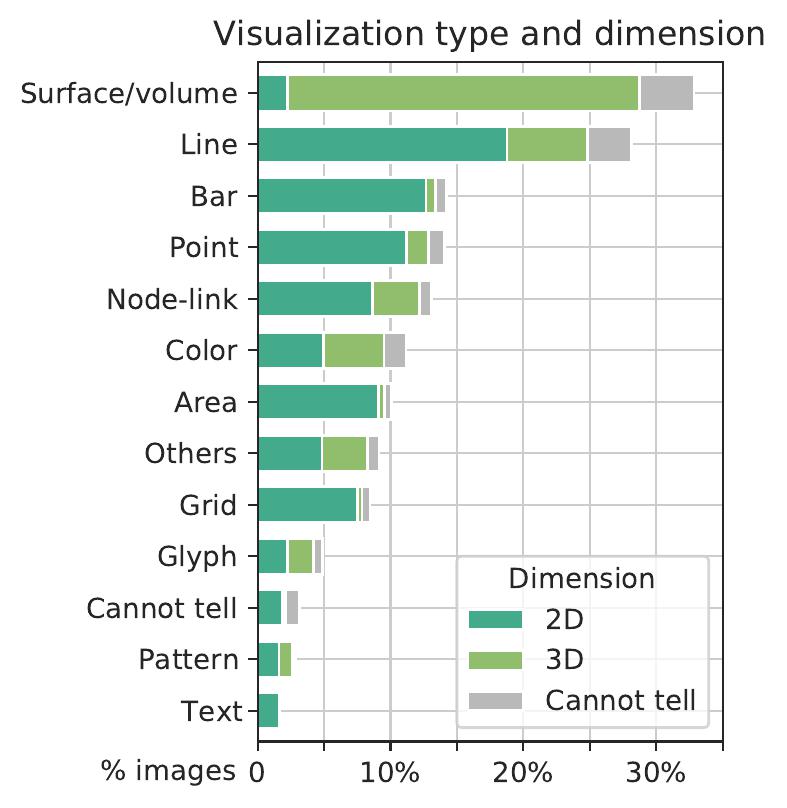}
    \caption{%
    The $x$-value represents the proportion of applied image codes in a category relative to the total \imagesfunction 
    visualization images (after excluding the pure schematic and GUI images).
    We can see that the most common visualization types were \vissurface and \visline.}
    \label{fig:typeDistribution}
\end{figure}

\begin{figure}[!bt]
    \centering
\subfloat[]{%
    \includegraphics[height=0.4\columnwidth]{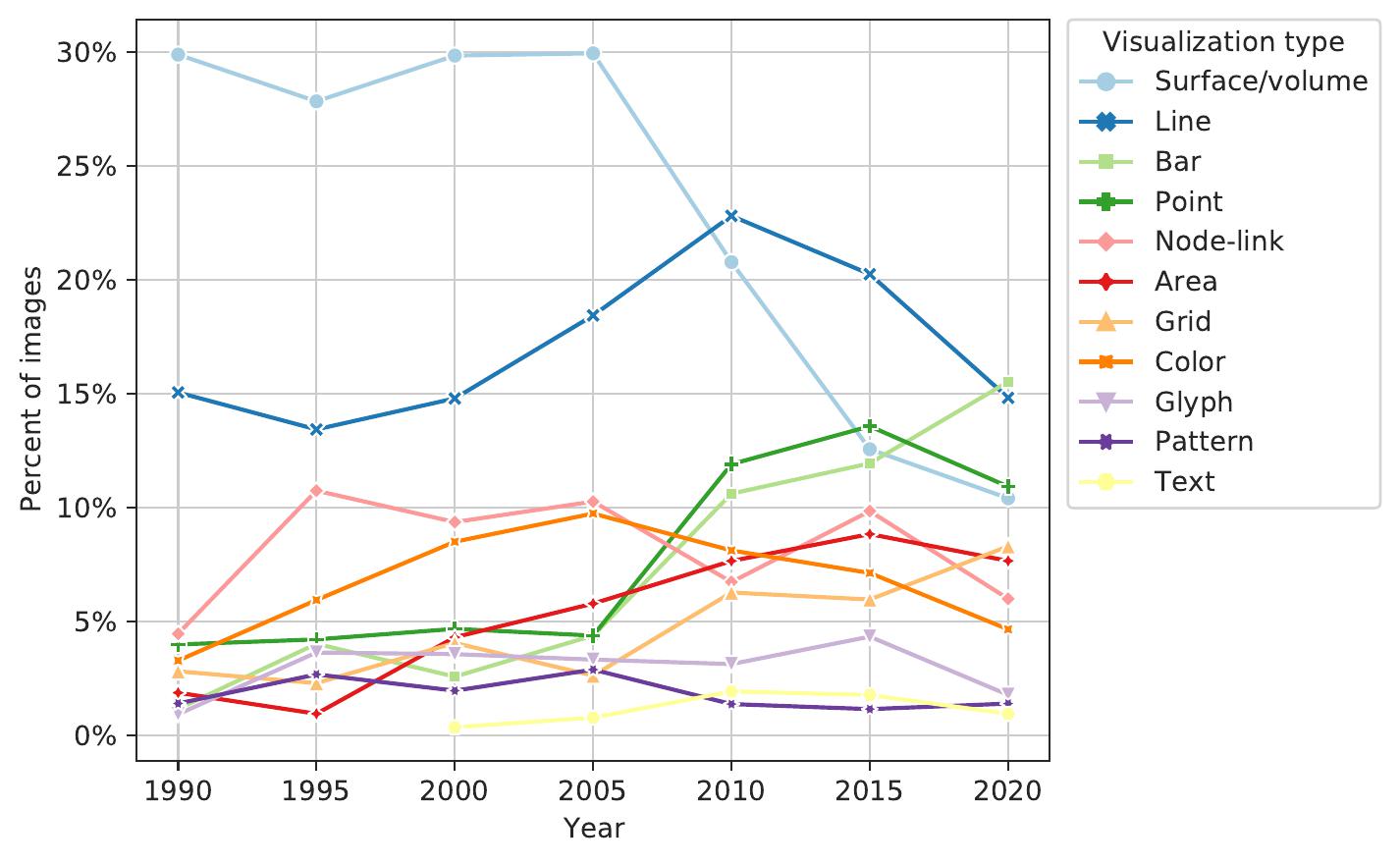}
    \label{fig:typeDistributionByYear}%
    }\hfill%
\subfloat[]{%
    \includegraphics[height=0.4\columnwidth]{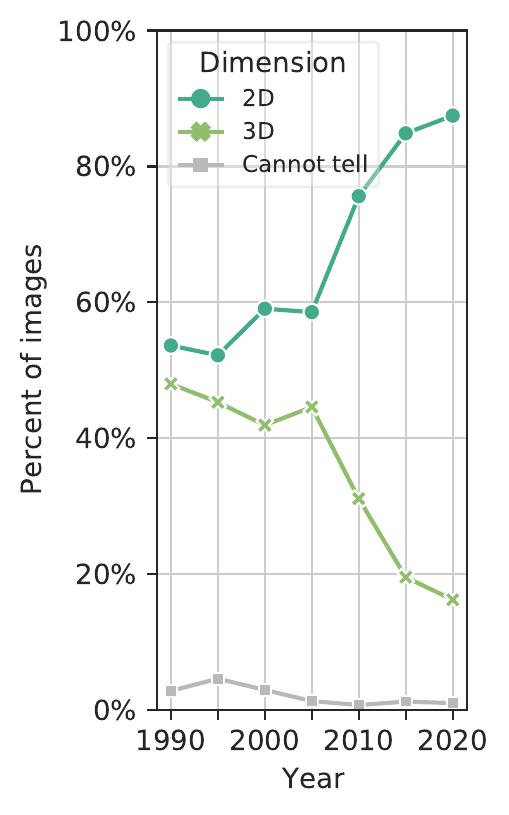}
    \label{fig:dimDistributionByYear}%
    }
    \caption{Temporal overview of the proportions of visualization types (a) and dimensions (b). We can see that \vissurface (Surface) become a lot less common after 2005 while at the same time \vispoint (Point) and \vislength (Bar) rise.
    }
    \label{fig:distributionByYear}
\end{figure}

\begin{figure}
\centering
\includegraphics[width=.6\columnwidth]{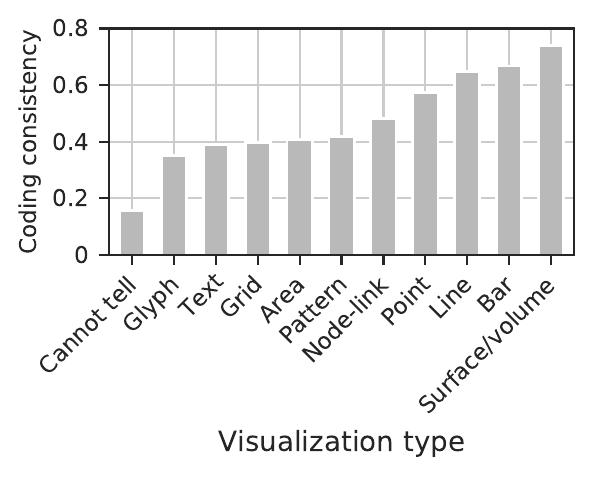}
\caption{The initial consistency of visualization type codes applied to images. We can see that the coders had least consistency related to \visglyph.}
\label{fig:consistency}
\end{figure}

In this section, we describe the results of our work after the completion of Phase 6 discussed in the previous section. We summarize the coding results based on the codes in each coding category.

Our coding process applied \totalfunctiontypelabels visualization type codes, 
{
\numguischematicCodeSingletons function codes,
} \totaldimlabels 2D/3D dimensionality codes, and \totalhardnesslabels difficulty codes to the \imagesfigSeven images. 
Many images included multiple types and dimensionality codes. \autoref{fig:typeDistribution}--left shows the distribution of visualization type
in relation to the difficulty ratings given by coders for each image. 
The right shows the distribution of dimensionalities per visualization type. \autoref{fig:distributionByYear} provides a historic overview of the spread of visualization type and dimensions  in a given year. 
\drpi{After our final
coding pass by the end of Phase~5, we calculated the consistency between the pair of coders that had coded each image. The resulting consistency depended strongly on the visualization type of the image (\autoref{fig:consistency}).}

\subsection{Visualization Types: Towards a Visualization Typology}
In this section, we describe canonical examples for each visualization type.
Later, in \autoref{sec:ambiguous}, we cover the main difficulties we encountered working with this typology. All images are referenced 
from left to right.  

\subsubsection{Generalized Bar Representations}
\label{sec:vc.bar}
%

As generalized bar charts we coded visualizations that represent data with straight or curved bars whose heights or lengths seemed proportional to represented data values.  Canonical examples of generalized bars are: (stacked/divided/regular) bar charts, histograms, radial bar charts, and donut charts. Generalized bar charts were the third most common visualization type among the images we coded and have gained in proportion in the later years we coded. Coders found them mostly easy to identify and the consistency between coders was among the highest at \barcon. 3D generalized bar charts were extremely rare and only sometimes appeared either in the early years we coded or more recently to visualize data on 3D surfaces such as a globe. 
Canonical examples of generalized bar charts taken from \cite{Weiss:2021:Revisited}, \cite{Ye:2021:ShuttleSpace}, \cite{Kim:2021:Githru}: 

\setlength{\pictureheight}{1.5cm}
\begin{figure}[h!]
\centering
\vspace{-1ex}
\includegraphics[height=\pictureheight]{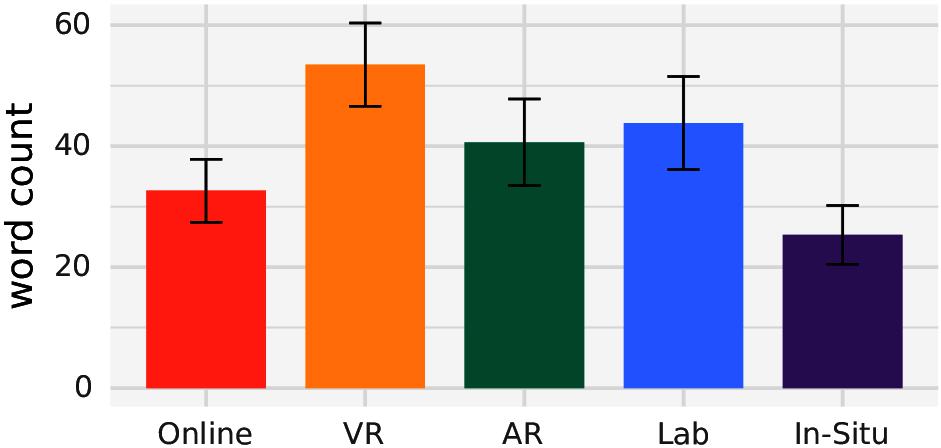}\hfill%
\includegraphics[height=\pictureheight]{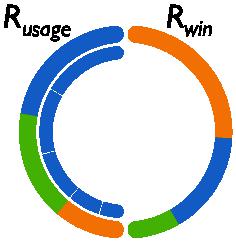}\hfill%
\includegraphics[height=\pictureheight]{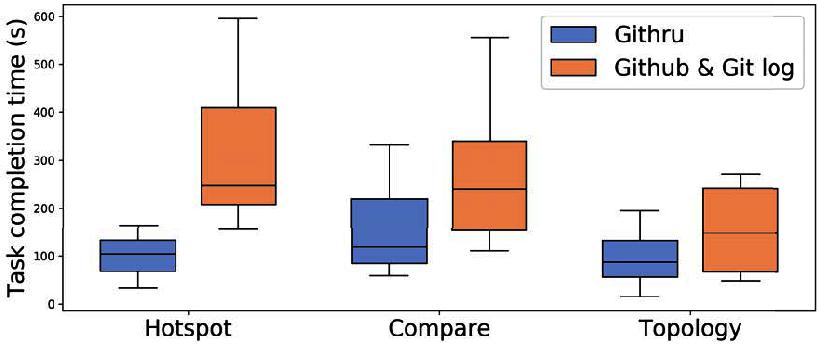}%
\vspace{-2ex}
\label{fig:bar}
\end{figure}

\subsubsection{Point-based Representations}
\label{sec:vc.point}
Point-based representations typically use dots or circles with a small 
radius to encode data, however, we
did not specify specific primitive shapes in our definition. Similar to Bertin~\cite{bertin1983semiology}, we considered point-based representation to encode point locations in a 2D or 3D space. Point marks could be small circles but also 3D spheres and sometimes other shapes like triangles, stars, etc. Canonical examples of visualization techniques of this type are scatterplots, (volumetric) point clouds, or dot plots. Point-based representations were the fourth most common visualization type according to our coding. Coders found identifying them slightly harder than length-based encodings and the overall consistency of coders was \pointcon. Surprisingly, we saw only a small percentage of 3D point-based representations, perhaps due to large amount of work on scatterplots or using scatterplot-like representations of, for example, dimensionality reduction or clustering results. 
Canonical examples of point-based representations taken from \cite{Pu:2000:AlgorithmVis},
\cite{xia2020smap},
\cite{ushizima2012augmented}:


\begin{figure}[h!]
\centering
\vspace{-1ex}
\includegraphics[height=\pictureheight]{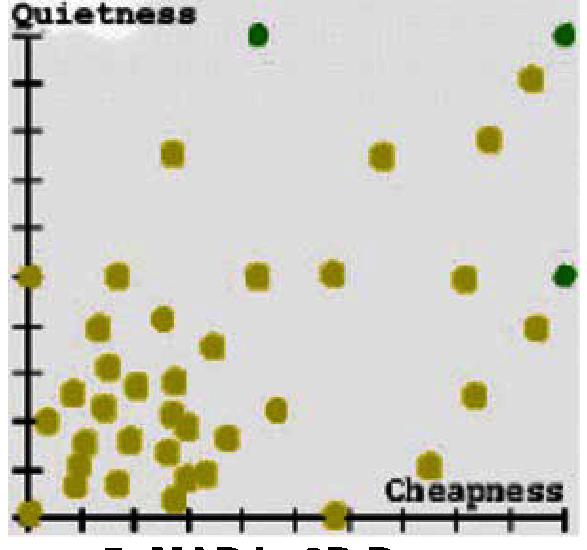}\hspace{10mm}%
\includegraphics[clip, trim={270 45 30 550}, height=\pictureheight]{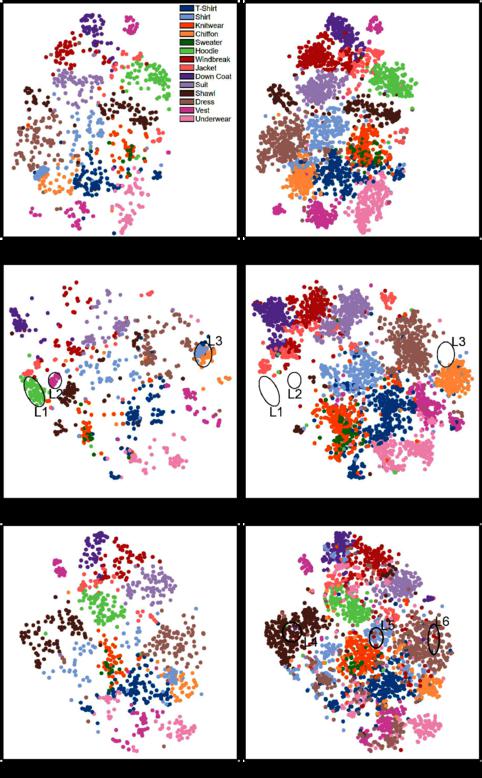}\hspace{7mm}%
\includegraphics[clip, trim={380 15 0 290}, height=\pictureheight]{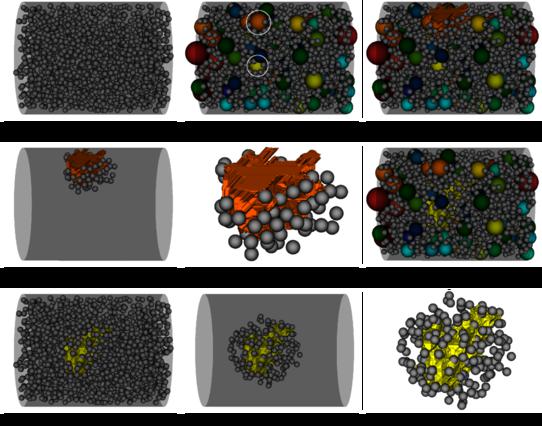}%
\vspace{-2ex}
\label{fig:points}
\end{figure}

\subsubsection{Line-based Representations}
\label{sec:vc.line}

Lines, edges, and curves were the second most common representations of data according to our codes.
Line-based visualization can depict surface features or data values. The lines and edges used in these visualizations could be straight or curved. 
Canonical line-based visualization techniques are line charts, parallel coordinates, contour lines, radar/spider charts, streamlines, or tensor field lines. We did not code lines that delineated areas as line-based representations. 

About a quarter (\lineprop) of line-based representations were rendered in 3D.  There was also a sizable proportion of \drpi{$12\%$} of
images where the coders could not tell whether a line chart was 3D or 2D due to a lack of clear depth cues. Most line charts, however, were the typical 2D line charts that most often represent temporal data. 
Images of canonical line-based representations taken from \cite{athawale2020direct}, \cite{stoll2005visualization},
\cite{verma2000flow}:

\begin{figure}[h!]
\centering
\vspace{-1ex}
\includegraphics[clip, trim={0 50 500 0}, height=\pictureheight]{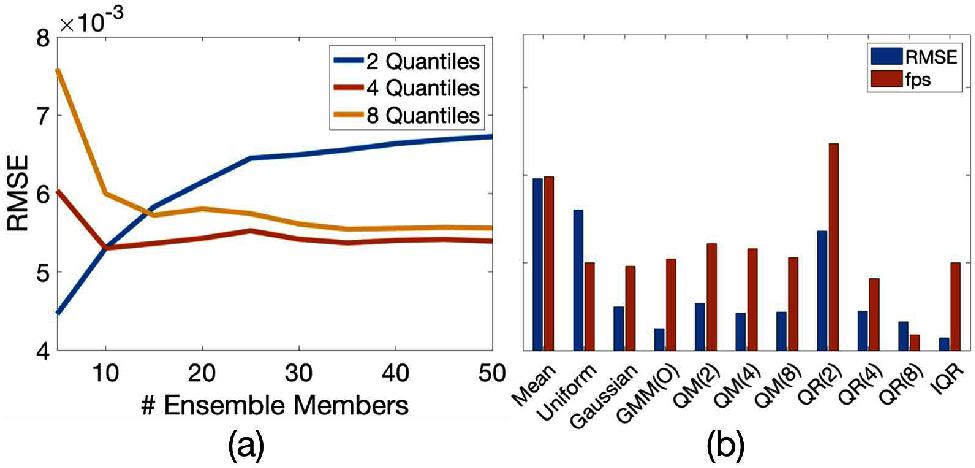}\hspace{7mm}%
\includegraphics[clip, trim={0 0 340 0}, height=\pictureheight]{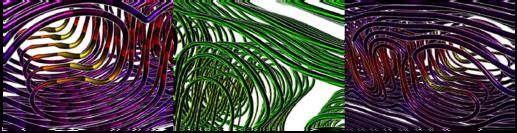}\hspace{7mm}%
\includegraphics[clip, trim={0 0 460 0}, height=\pictureheight]{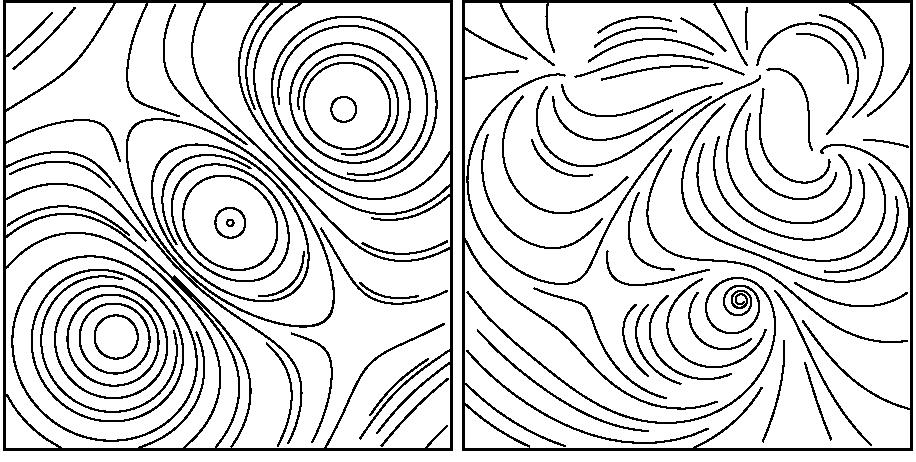}%
\vspace{-2ex}
\label{fig:line}
\end{figure}

\subsubsection{Node-link Trees/Graphs, Networks, Meshes}
\label{sec:vc.nodelink}

Representations of this type depict points and explicit connections to convey relationships between data values. Node-link relationships can be found in trees, graphs, networks, and meshes. Node positions can be given, \eg, geospatial locations, or be coded in the data (\eg, projections). Connections can be continuous, \eg, a Reeb graph, as the topological structure is given by showing continuous functions in space, or discrete, \eg, edges in a tree. Representations of this type were the 5\textsuperscript{th} most common representation type and their representation has stayed relatively stable at roughly 5--10\% of images per year. Most codes were applied to 2D images, but \drpi{$27\%$} of the codes in this category belonged to images in 3D. Canonical \visnodelink representations taken from \cite{phan2005flow},
\cite{yoghourdjian2020scalability},
\cite{bhaniramka2000isosurfacing}:

\begin{figure}[h!]
\centering
\vspace{-1ex}
\includegraphics[height=\pictureheight]{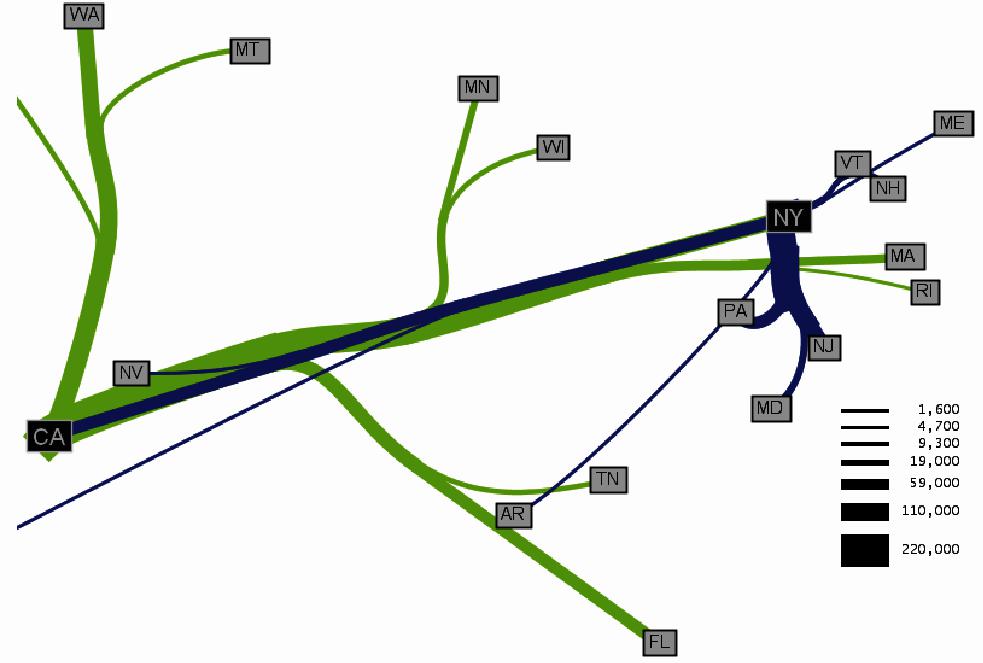}\hspace{7mm}%
\includegraphics[clip, trim={900 1306 300 650}, height=\pictureheight]{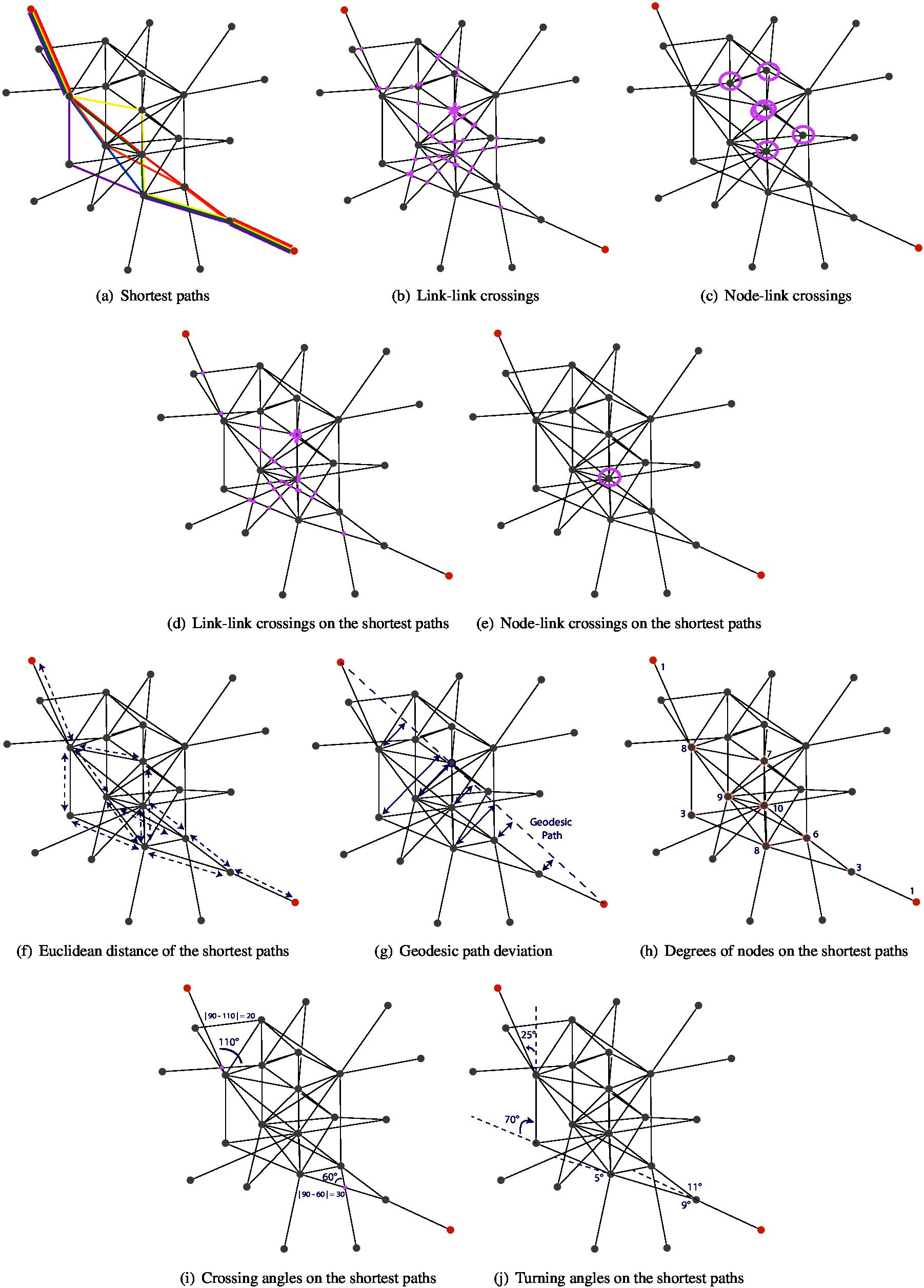}\hspace{7mm}
\includegraphics[clip, trim={500 30 0 0}, height=\pictureheight]{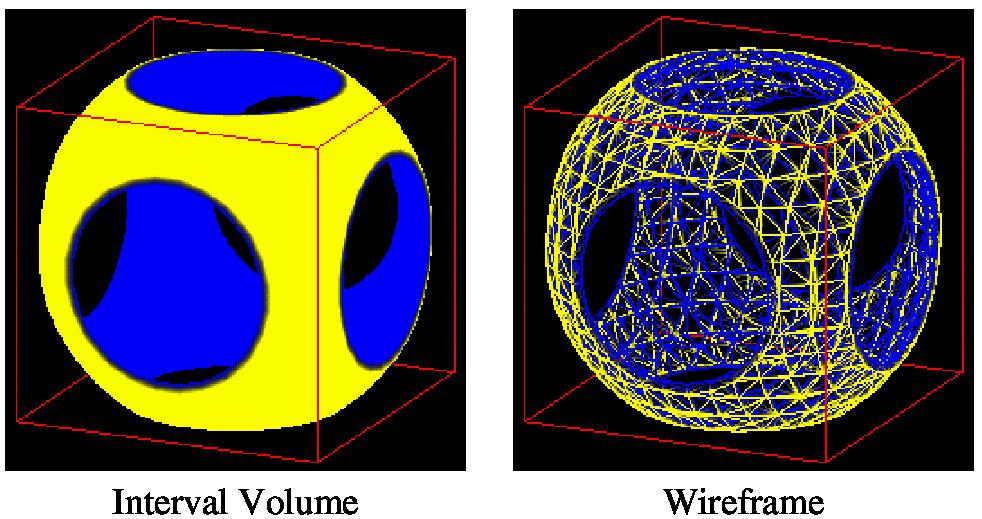}%
\vspace{-2ex}
\label{fig:nodelinkMeshes}
\end{figure}



%
%



\subsubsection{Generalized Area Representations}
\label{sec:vc.area}
Generalized area charts are representations with a focus on areas of 2D space or 2D surfaces, including sub-sets of these surfaces. Areas can be geographical regions or polygons whose size or shape represents abstract data. Areas often feature explicit boundaries and, within, are filled with categorical colors or use  contrast in luminance and shading to encode attributes of the areas. Common examples of generalized area charts are regular area charts, treemaps, cartograms, choropleth maps, pie charts, or violin plots.  Generalized area charts were the 6\textsuperscript{th} most common type of representation type in the images we coded. Most areas were part of surfaces rendered in 2D. Over the years the proportion of images with areas increased by 1--2\% every 5 years up to just under 10\% in recent years. 
Canonical examples of \visarea taken from \cite{bu2020sinestream},
\cite{shneiderman2001ordered},
\cite{rheingans1995interactive}:

\begin{figure}[h!]
\centering
\vspace{-1ex}
\includegraphics[clip, trim={60 700 1120 70}, height=\pictureheight]{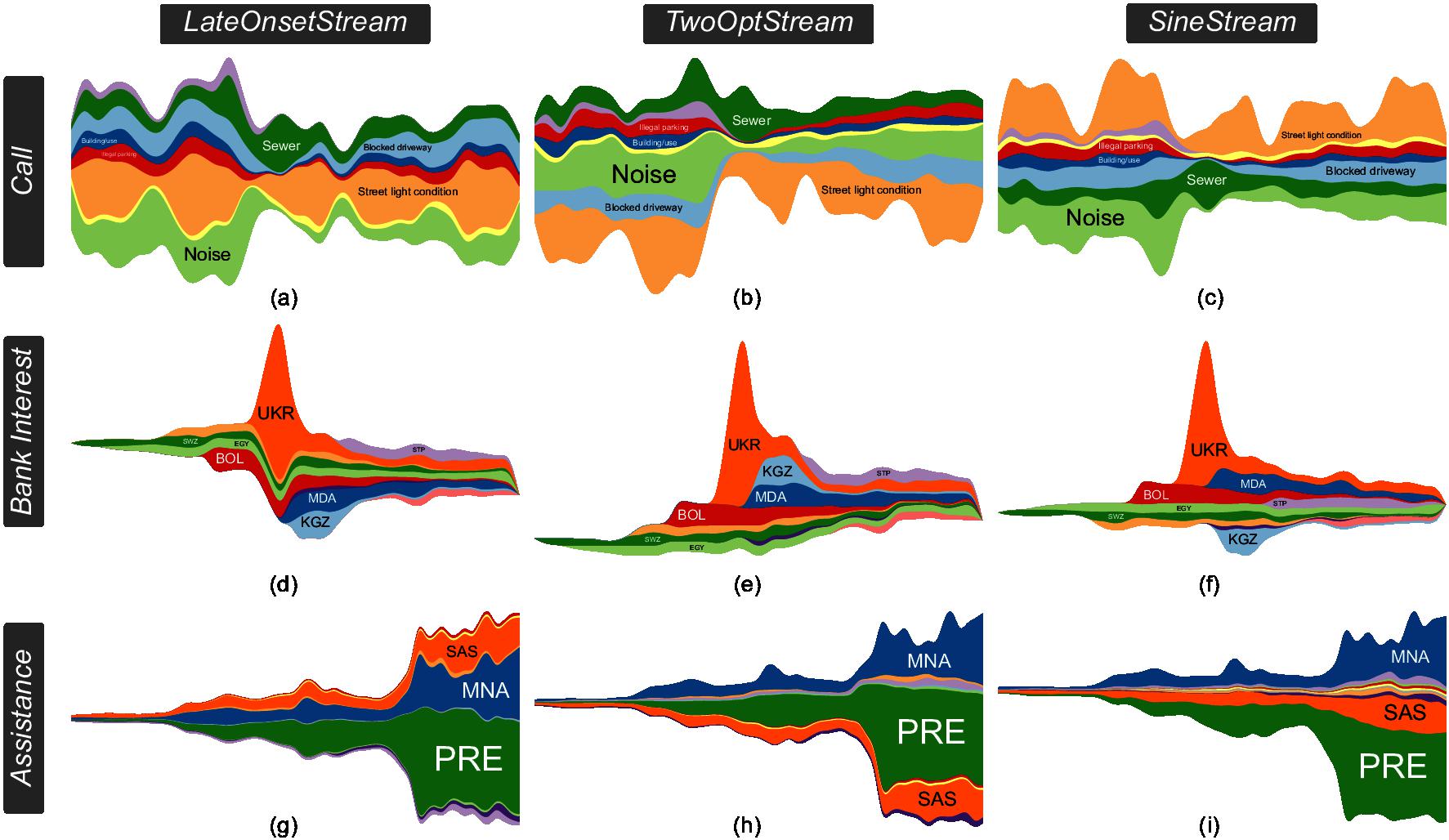}\hfill%
\includegraphics[height=\pictureheight]{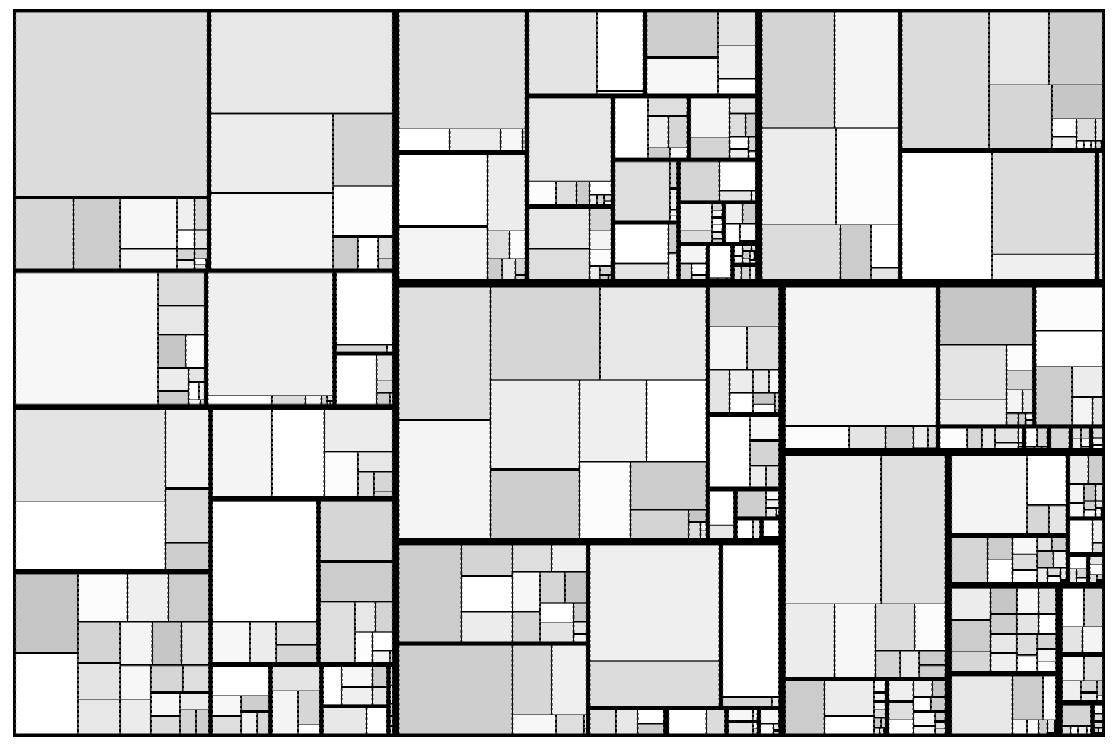}\hfill%
\includegraphics[height=\pictureheight]{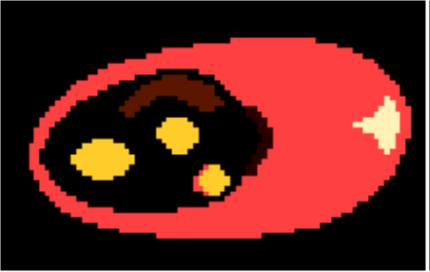}%
\vspace{-2ex}
\label{fig:areas}
\end{figure}

\subsubsection{Surface-based Representations and Volumes}
\label{sec:vc.surfacevolume}
Surfaces and volumes represent the inner and/or outer features and/or boundaries of a continuous spatial phenomenon or object in 3D physical space, 4D space-time, or slices thereof.  Surfaces typically represent the inner and outer boundaries of a given 3D scalar field,  \eg, isosurfaces~\cite{gregorski:adaptive},  or integral surfaces in vector fields,  \eg, stream ribbons and stream surfaces~\cite{edmunds:surface}. Volume rendering is a set of dedicated techniques that depict sub-sets of volume data, usually with an element of thickness (as opposed to infinitely thin surfaces)~\cite{hansen:visualization}. 
Frequent characteristics of images in this category include:
the use of semi-transparency,
the application of lighting and shading techniques,
perspective or parallel projection,
and the addition of other 3D and depth cues.
Volume rendering has been one of the most important areas of Visualization in the early years of the conference \cite{isenberg2016visualization}. As such, it is perhaps not surprising that surface and volume representations were the most common techniques in our dataset. It is also the only representation technique with primarily 3D renderings. The few 2D renderings we found included slices of volumes such as X-ray or CT slices. Similar to what we saw in prior work on keywords \cite{isenberg2016visualization}, the number of surface and volume renderings has decreased drastically since 2005 from around 25--30\% of all images to currently around 15\%.
Canonical examples of surface and volume-based representations taken from \cite{hummel2010iris}, \cite{tikhonova2010visualization},
\cite{nguyen2021modeling}:


\begin{figure}[h!]
\centering
\vspace{-1ex}
\includegraphics[clip, trim={0 40 560 0},height=0.18\columnwidth]{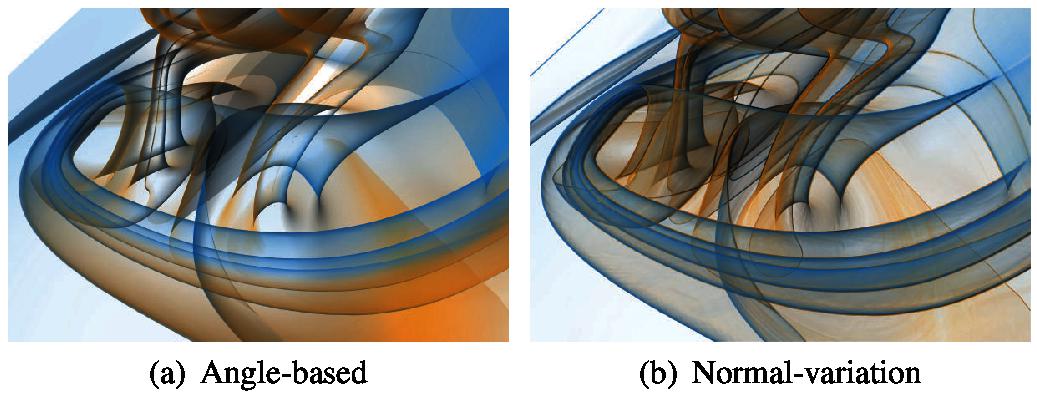}\hspace{7mm}%
\includegraphics[clip, trim={400 240 1200 0},height=0.18\columnwidth]{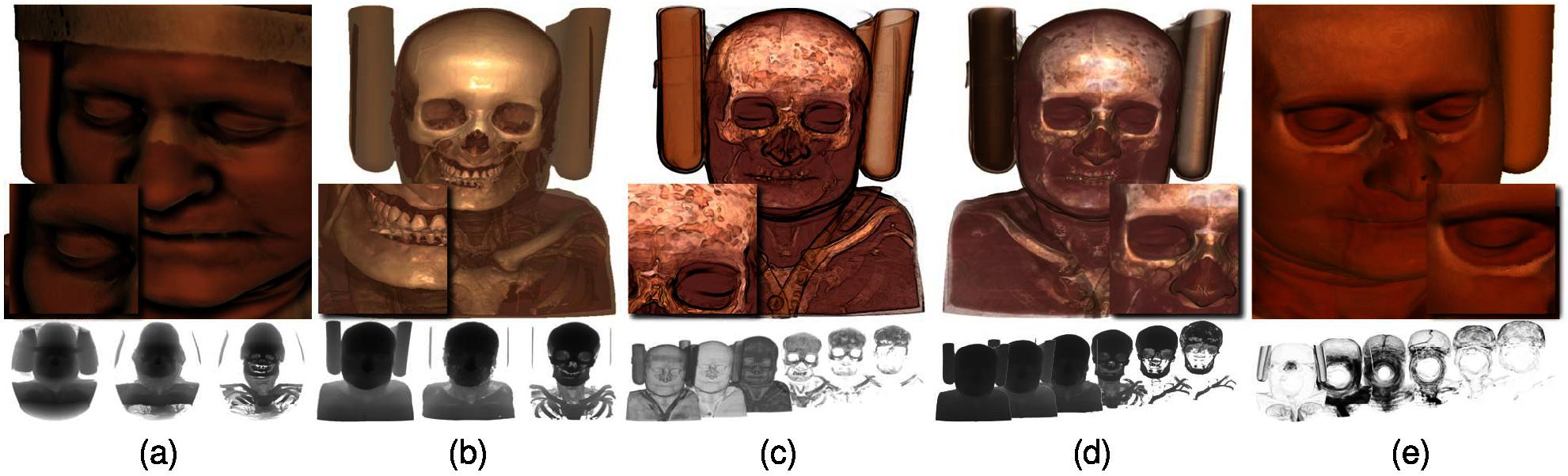}\hspace{5mm}%
\includegraphics[clip, trim={600 0 0 0}, height=0.17\columnwidth]{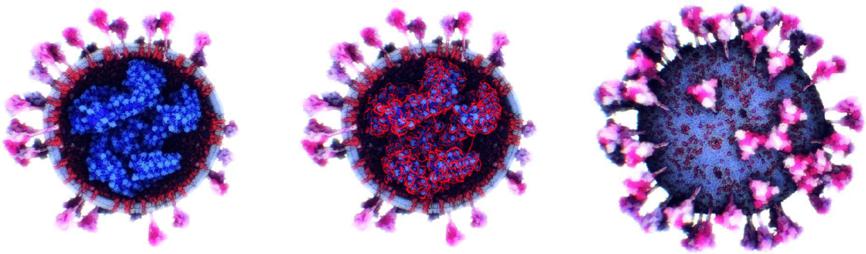}%
\vspace{-2ex}
\label{fig:surfaceVolume}
\end{figure}

%
%

\subsubsection{Generalized Matrix or Grid}
\label{sec:vc.grid}
Generalized matrices and grids are representations that position data in a \textit{discrete} grid structure.  
The grid can vary in resolution, is typically rectilinear but can use other shapes such as hexagonal grids etc.
This representation type includes figures where the underlying grid is part of the data structure, but does not include figures where the underlying grid is merely used as a convenient arrangement of sub-sets of the data (as in small-multiples and scatterplot matrices). Other elements such as color or glyphs can appear at these discrete grid positions (\eg, grid-based vector field visualization).
Under 10\% of all images contained generalized matrices/grids and that consistently across the years. Of these, \drpi{89\%} were 2D grids/matrices and \drpi{4.2\%} were 3D (the rest was ``I cannot tell''). Common visualization techniques in this representation type are discrete heatmaps, scarf/strip plots, space-time cubes, or matrix-based network visualizations. 
Canonical examples of this visualization type taken from
\cite{ingram2010dimstiller},
\cite{Nielson:1991:TAD},
\cite{garcke2000continuous},
\cite{meulemans2020simple}:

\begin{figure}[h!]
\centering
\vspace{-1ex}
\includegraphics[clip, trim={500 0 30 30}, height=0.2\columnwidth]{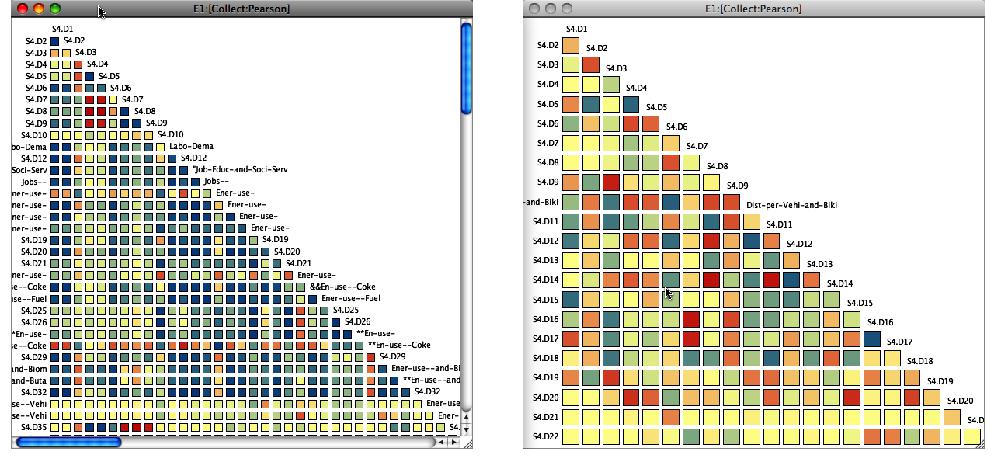}\hfill%
\includegraphics[clip, trim={230 160 240 80},height=0.18\columnwidth]{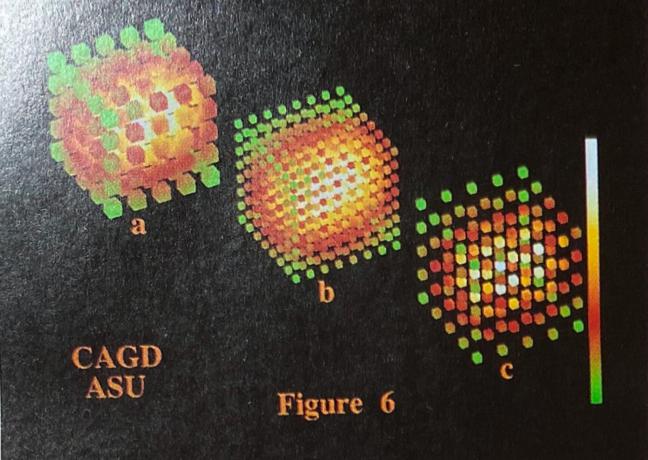}\hfill%
\includegraphics[width=0.2\columnwidth]{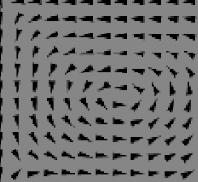}\hspace{1mm}%
\includegraphics[clip, trim={500 150 1000 150}, height=0.2\columnwidth]{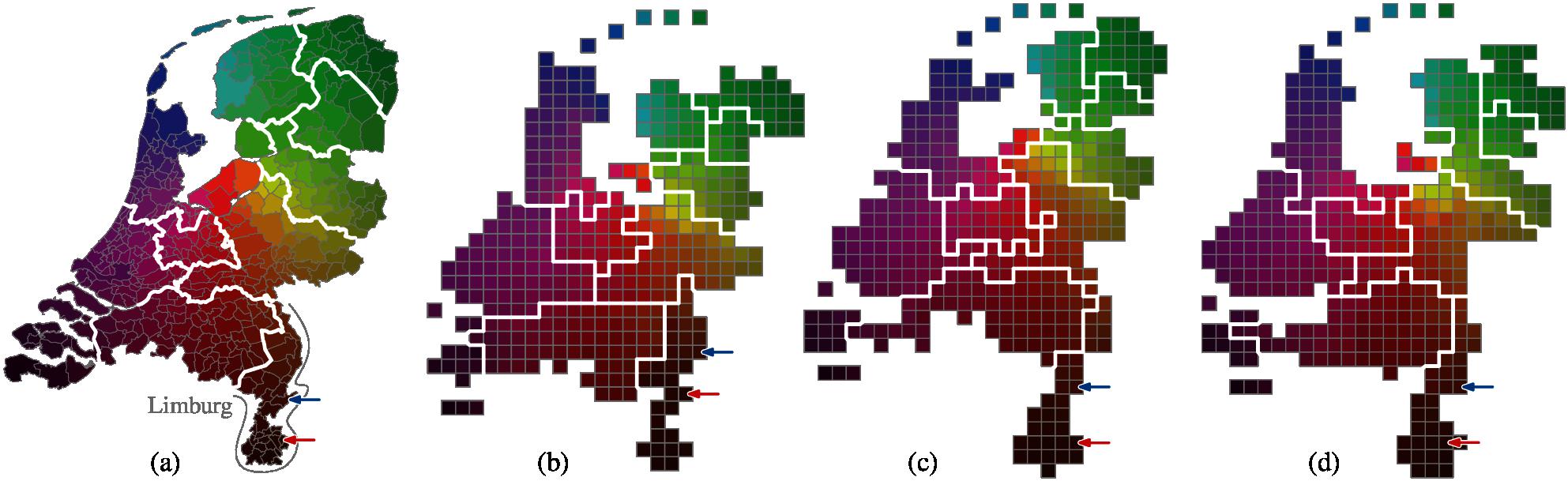}%
\vspace{-2ex}
\label{fig:grid}
\end{figure}

\subsubsection{Continuous Pattern-based Representations}
\label{sec:vc.pattern}

In general, continuous pattern-based representations incorporate images that focus on representing  continuous data variation along planes and surfaces (akin to a ``texture''). This category differentiates itself from  \viscolor (see next) by using repetitive patterns or structures in the texture mapped to data. Frequent characteristics of representations of this type are smooth, high-resolution, and highly detailed variations/changes across the data. Representations of simulated flow are common.  Typical visualization techniques include: Line Integral Convolution (LIC), Spot Noise, Image-Space Advection (ISA), or Image-Based Flow Visualization (IBFV). Continuous pattern-based representations were among the most rare types in our coding with under \drpi{\patternprop} of all images in our dataset. Most patterns were used on surfaces represented in 2D, but some also applied to 3D. 
Canonical examples of this visualization type taken from
\cite{garcke2000continuous},
\cite{van2003image},
\cite{jobard2000hardware}:


\begin{figure}[!ht]
\centering
\vspace{-1ex}
\includegraphics[height=\pictureheight]{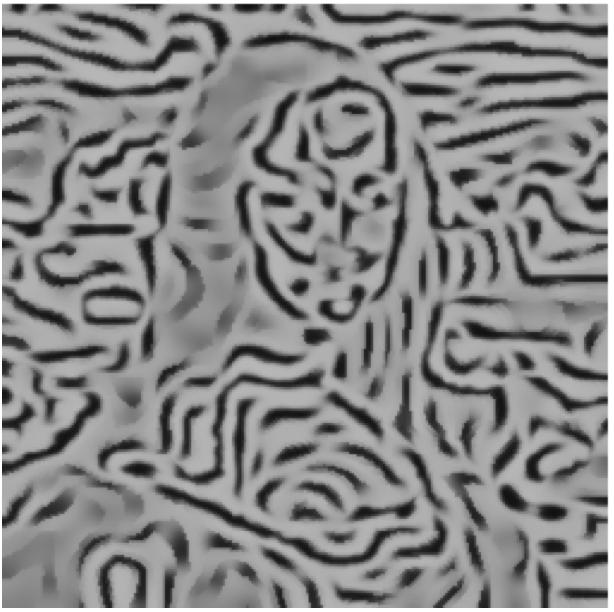}\hspace{7mm}%
\includegraphics[clip, trim={0 0 500 0}, height=\pictureheight]{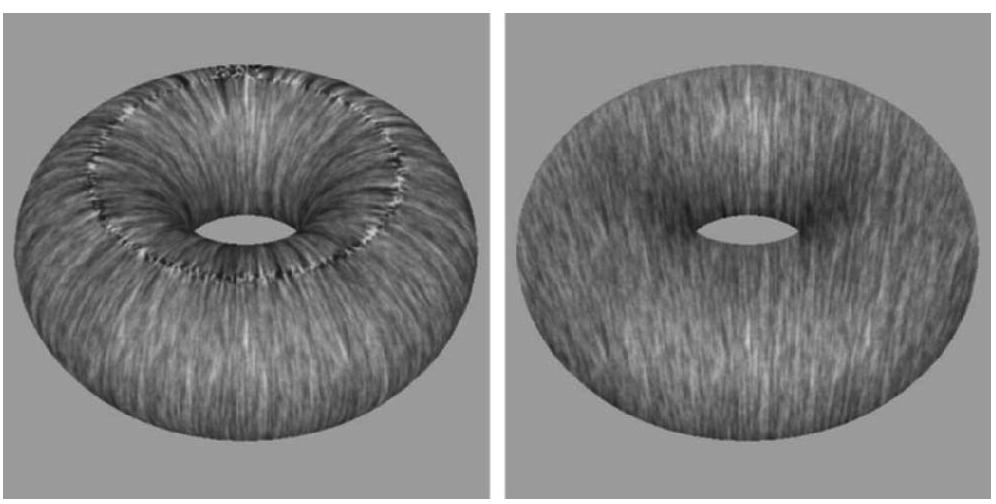}\hspace{6mm}%
\includegraphics[clip, trim={640 0 0 0}, height=\pictureheight]{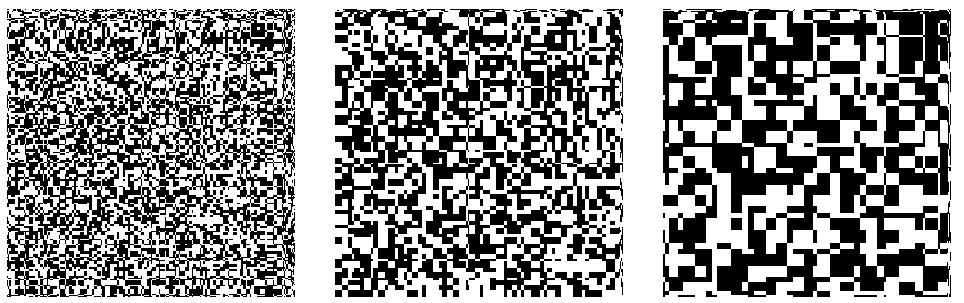}%
\vspace{-2ex}
\label{fig:pattern}
\end{figure}





\subsubsection{Continuous Color-based Representations}
\label{sec:vc:color}
Continuous Color-based Representations use a primary color (hue, brightness, and/or saturation) encoding across a continuous surface or volume. We added this visualization type late during the verification phase (Phase~6 in \autoref{sec:process}) because many of our discussions arose around how to code heatmaps (\eg, right-most image below). Discrete heatmaps were clearly of Matrix/Grid type, but other similar color-based encodings applied in a continuous way were not. 
The main characteristics of continuous color-based representations are the prominent encoding of data through color at high resolution down to the individual pixel level and
smooth transitions between varying colors. 

A prominent technique in this category are representations featuring a transfer function. 
Conceptually, a transfer function is a colormap with an added opacity encoding.  Other examples include continuous heatmaps and pixel-based encodings.  Color-based encodings are less common in our data but this can be partly attributed to the fact that the code only appeared in the final verification phase and that some images might have been missed, this is why we do not provide a consistency score for this code. Most continuous color-based encodings were applied to surfaces rendered in 2D. 
Canonical examples of this visualization type from left to right taken from
\cite{demiralp2009coloring},
\cite{feng2020topology},
\cite{wang2008importance},
\cite{fisher2007hotmap}:


\begin{figure}[h!]
\centering
\vspace{-1ex}
\includegraphics[clip, trim={350 60 0 20}, height=\pictureheight]{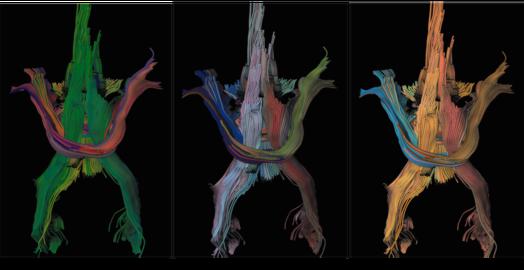}\hfill
\includegraphics[clip, trim={50 20 780 100},height=\pictureheight]{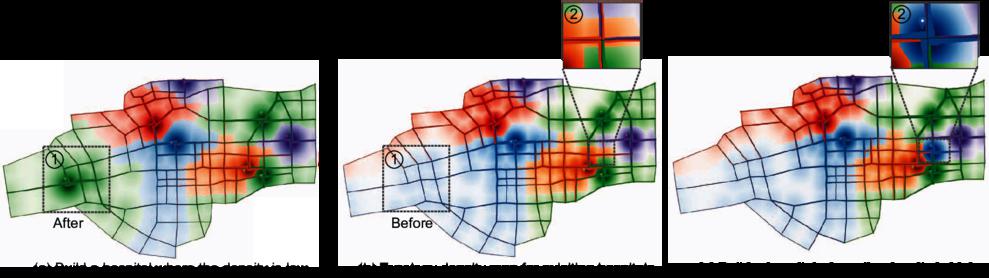}\hfill
\includegraphics[clip, trim={500 0 0 0},height=\pictureheight]{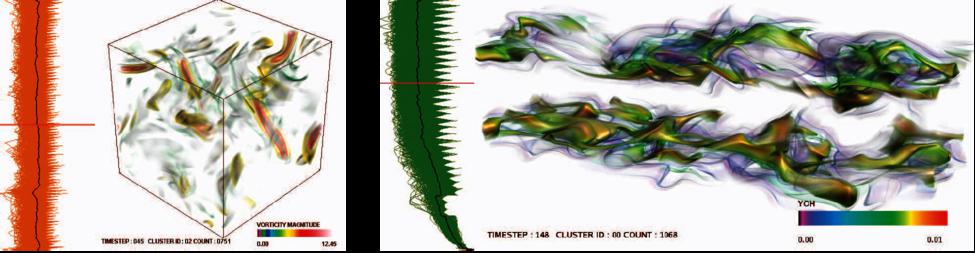}\hfill
\includegraphics[clip, trim={50 80 50 70}, height=\pictureheight]{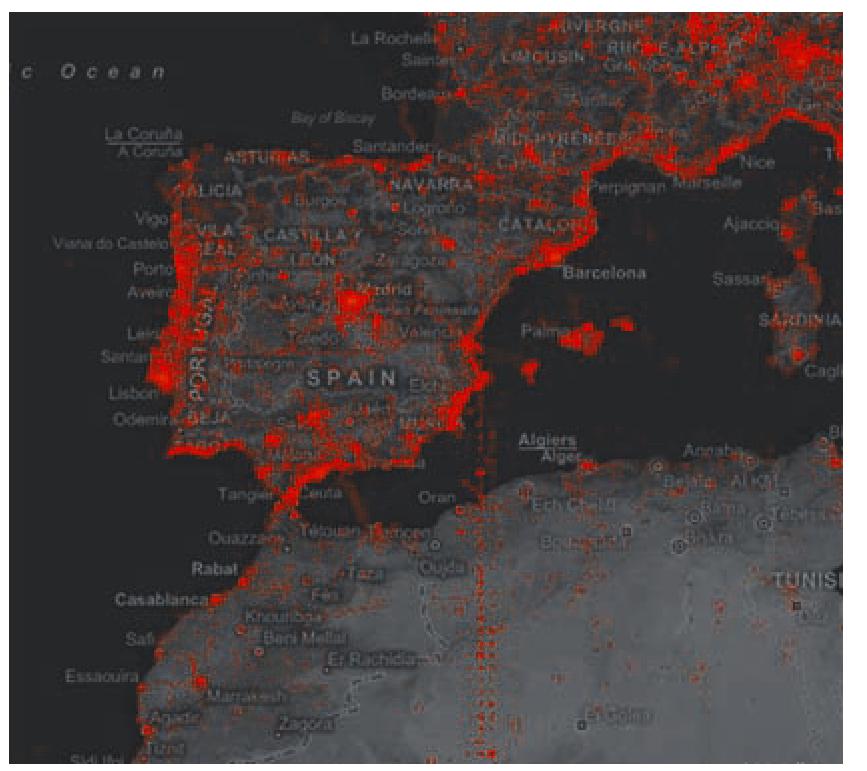}%
\vspace{-2ex}
\label{fig:color}
\end{figure}

\subsubsection{Glyph-based Representations}
\label{sec:vc.glyph}
We coded glyphs when we saw multiple small independent visual representations that depicted multiple attributes (dimensions) of a data record. 
These glyphs often used multiple geometric primitives to encode data.
When multiple properties of a single mark encoded data, we also considered them as glyphs especially when we saw multiple glyphs displayed for comparison and/or with meaningful placement in space. 
For example, 3D glyphs, often were made up of one mark such as a small 3D cuboid  where height, width, and depth could encode different data dimensions. In much of the visualization literature these marks would be named a glyph and we retained this usage of the term. Common glyph-based techniques in this category were star glyphs \inlinevis{-2pt}{1.1em}{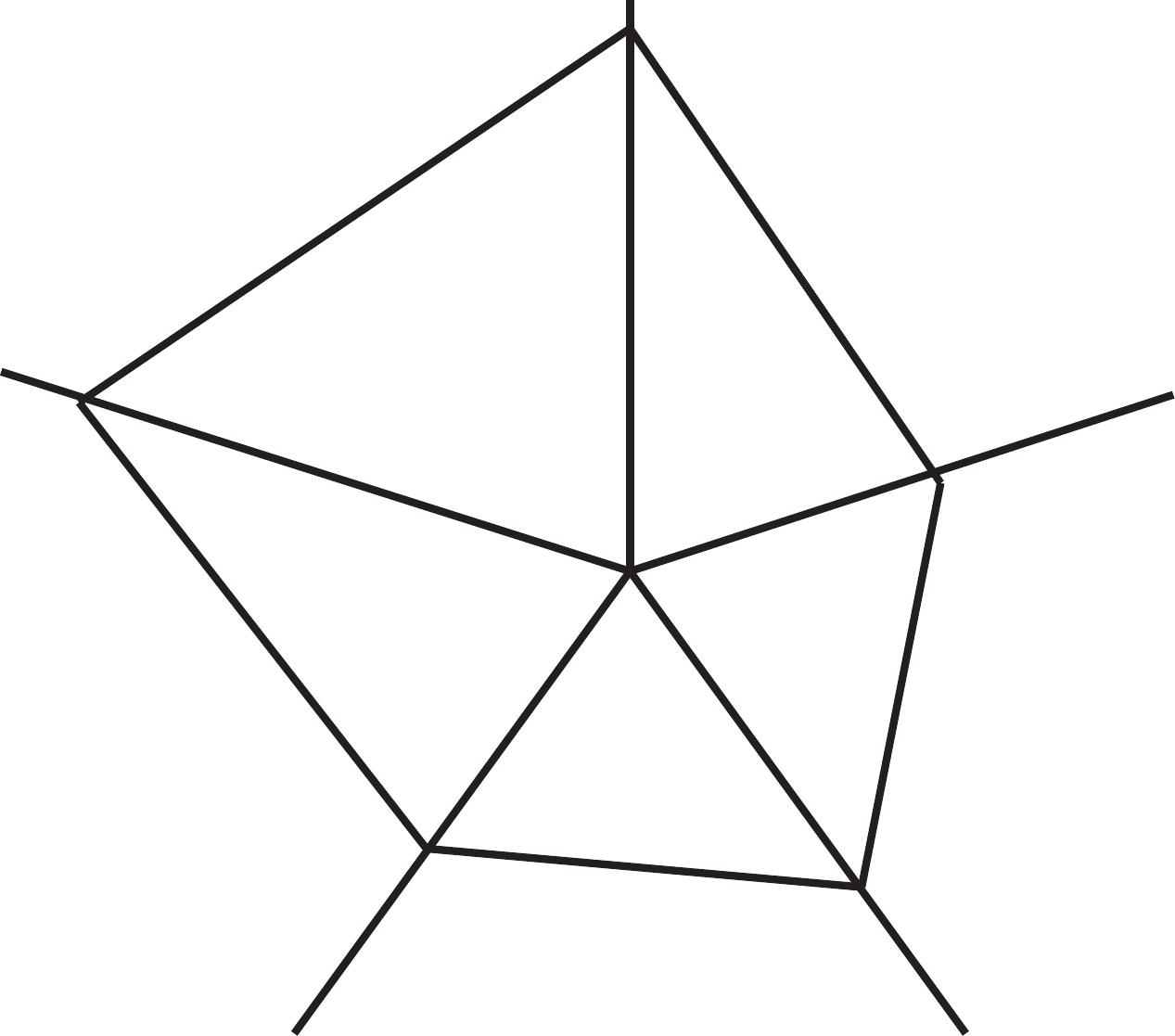}, 3D tensor glyphs \inlinevis{-2pt}{1.1em}{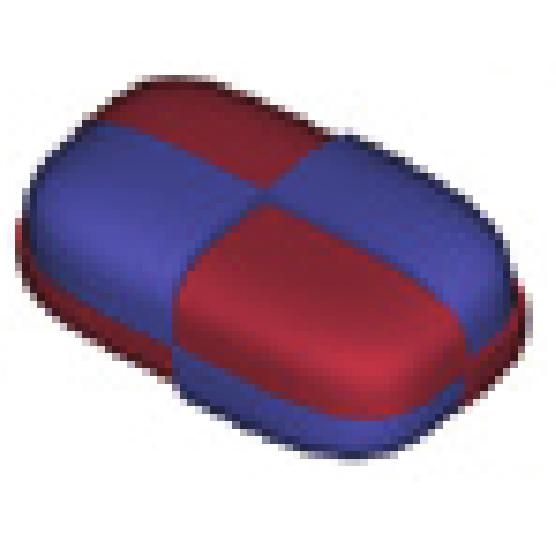}, Chernoff faces \inlinevis{-2pt}{1.1em}{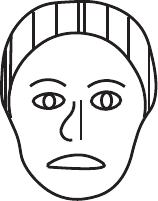}, or vector field arrows \inlinevis{-2pt}{1.1em}{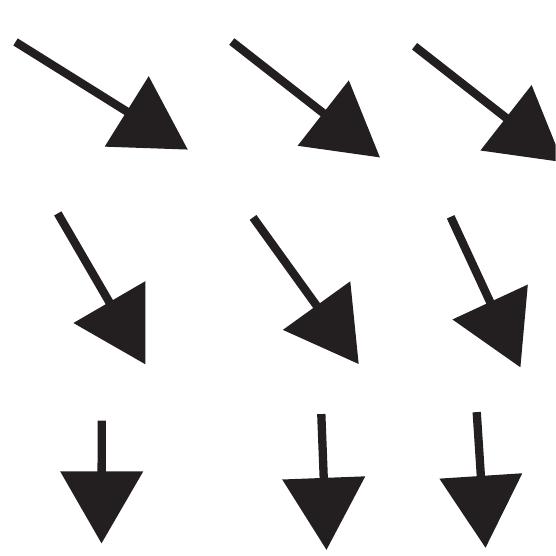}. Overall, glyph-based encodings were not particularly frequent ($<$5\% of all images contained glyphs) and we saw only slightly more glyphs rendered in 2D than 3D. Glyphs, however, were difficult to identify and our consistency for this visualization type was initially only \glyphcon. 
Canonical examples of this visualization type taken from
\cite{fanea2005interactive},
\cite{hlawatsch2011flow},
\cite{hlawitschka2005hot},
\cite{garcke2000continuous}:


\begin{figure}[!h]
\centering
\vspace{-1ex}
\includegraphics[clip, trim={0 0 800 0}, height=\pictureheight]{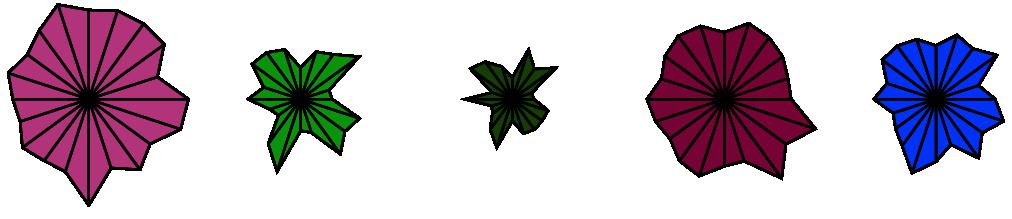}\hfill%
\includegraphics[clip, trim={473 40 0 0}, height=\pictureheight]{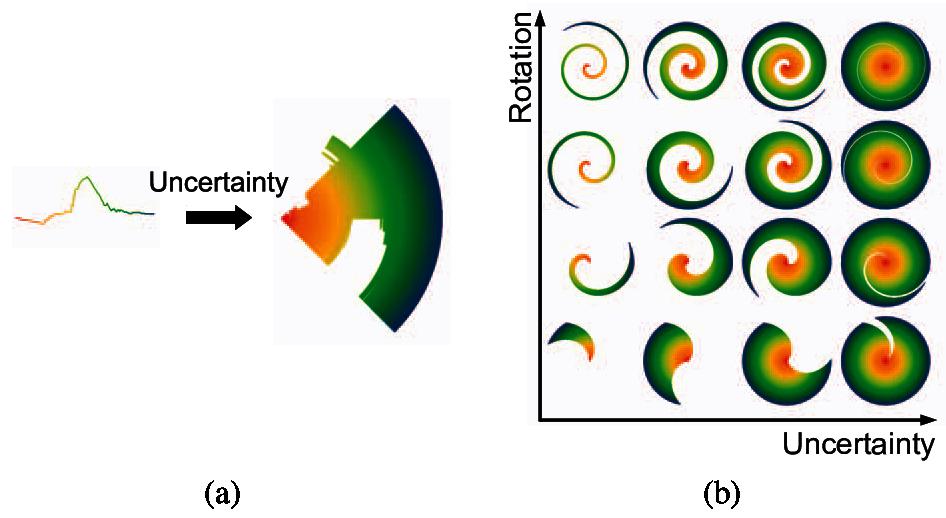}\hfill%
\includegraphics[height=0.5\pictureheight]{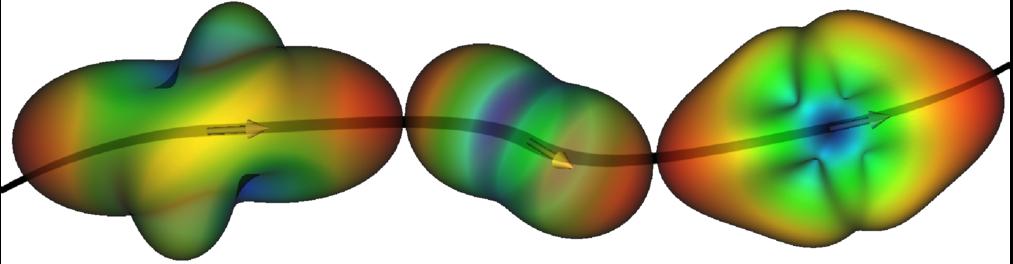}\hfill
\includegraphics[height=\pictureheight]{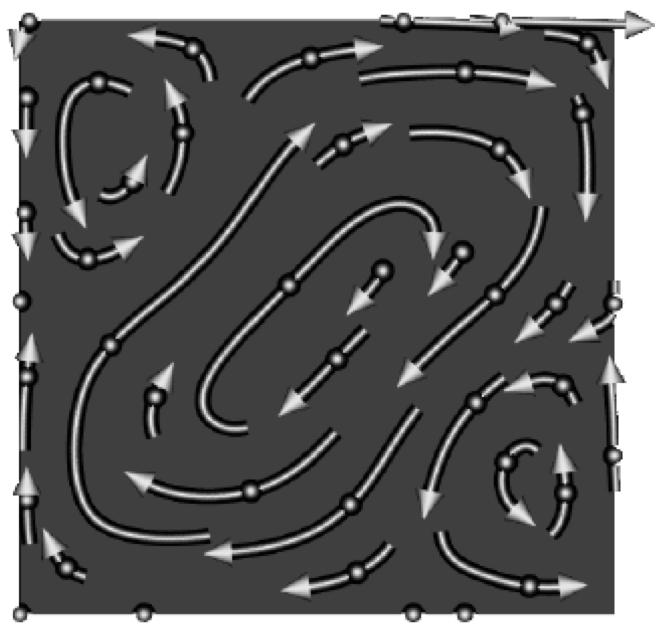}%
\vspace{-2ex}
\label{fig:glyphs}
\end{figure}

\subsubsection{Text-based Representations}
\label{sec:vc.text}
Text-based representations encode data (usually text itself) using varying properties of letters and words such as font size, color, width, style, or type.
Common visualization techniques for this representation type are tag clouds, word trees, or typomaps. We did not code images where text was used only for labeling and annotation or where text was the underlying data source but the representation did not use text properties to encode the text. Text-based representations were the most rare in our coding. All text-based representations were rendered in 2D. The initial coding consistency was low at \textcon primarily because some coders initially also coded representation of text as a data source. 
Example images for this type taken from \cite{lee2010sparkclouds},
\cite{wang2017edwordle},
\cite{afzal2012spatial}:


\begin{figure}[h!]
\centering
\vspace{-1ex}
\includegraphics[clip, trim={100 200 200 200},height=\pictureheight]{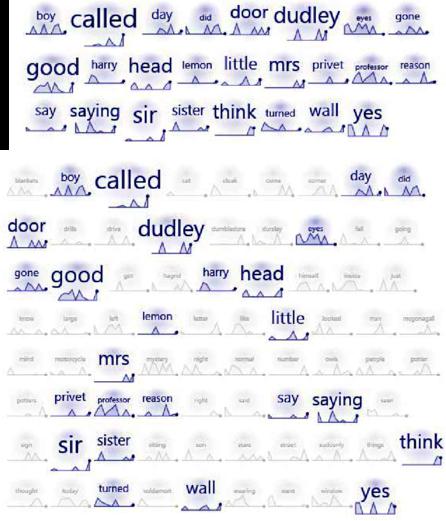}\hspace{7mm}%
\includegraphics[clip, trim={0 50 1500 0}, height=\pictureheight]{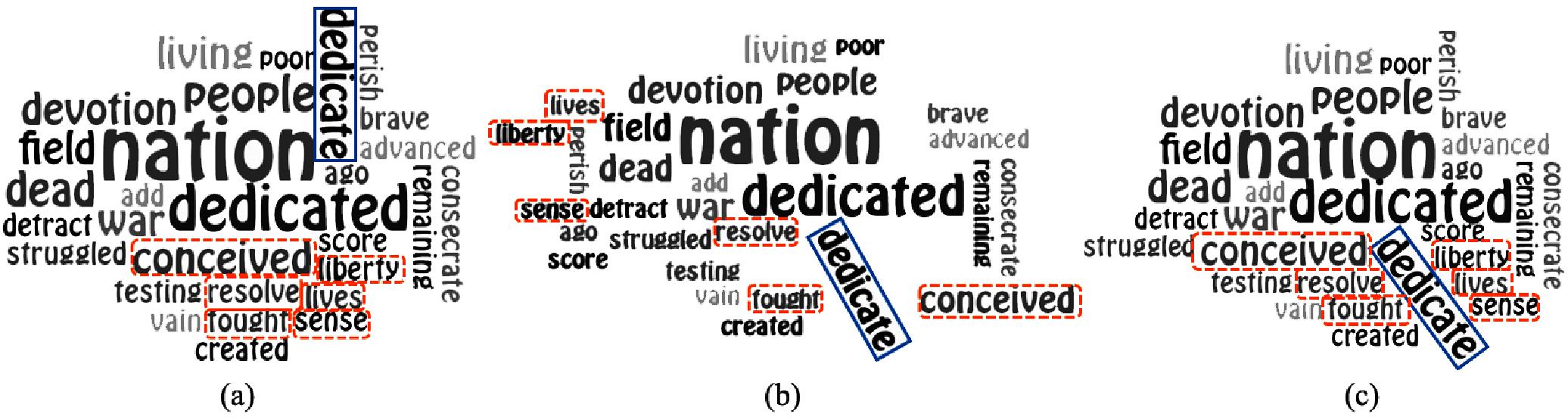}\hspace{7mm}%
\includegraphics[height=\pictureheight]{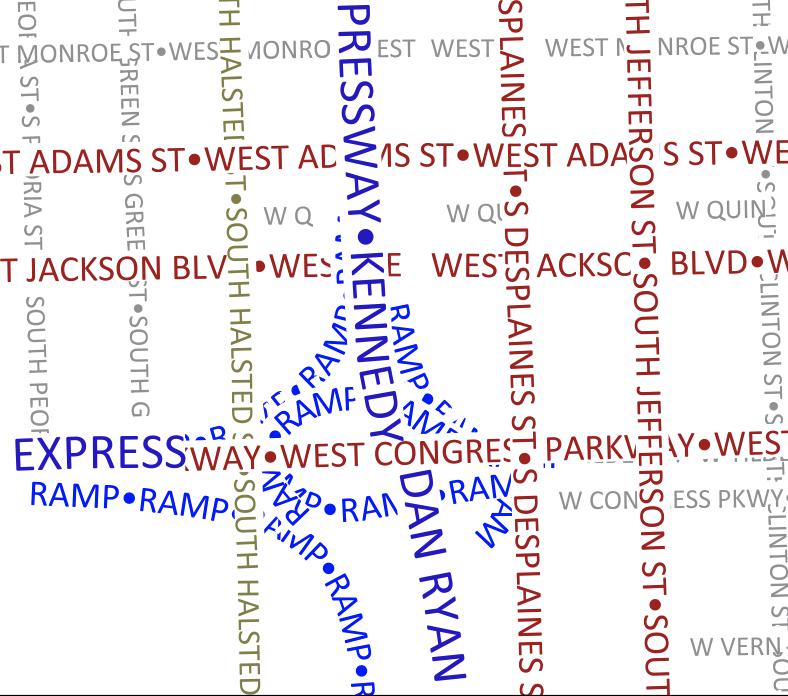}%
\vspace{-2ex}
\label{fig:text}
\end{figure}





\subsection{Visualization Functions}

All images tagged with a visualization type code were implicitly assigned the function to ``showcase a visualization technique.'' This function was by far the most common. In addition to this implicit function we coded screenshots or images of graphical user interfaces (GUIs) and schematic representations. For images with both of these functions we did not assign additional individual visualization types. 

\subsubsection{GUI (Screenshots)/User Interface Depiction}
\label{sec:vc.gui}
Images tagged with this code were either screenshots or photos of a system interface.  GUIs required the presence of window components or other UI widgets, such as buttons, sliders, boxes, scroll bars, pointers (\eg, the hand cursor showing interaction), etc. 
Non-WIMP interfaces (\eg, for VR or touch-based applications) were indicated by, \eg, a hand\discretionary{/}{}{/}finger touching a surface or clearly visible interface hardware such as a tablet, a tabletop display, or other types of hardware.
{There are \numguiOrg GUI images in total representing about $12\%$ of all images. The proportions of the GUI images over time are also relatively stable over the years. 
Only about $13\%$ of these are 3D.
Coders found identifying them was easy and the overall consistency of coders was \guicon.}
Canonical examples for this type taken from \cite{bergman1995rule}, 
\cite{dasu2020sea},
\cite{yoghourdjian2020scalability}:


\begin{figure}[h!]
\centering
\vspace{-1ex}
\includegraphics[height=\pictureheight]{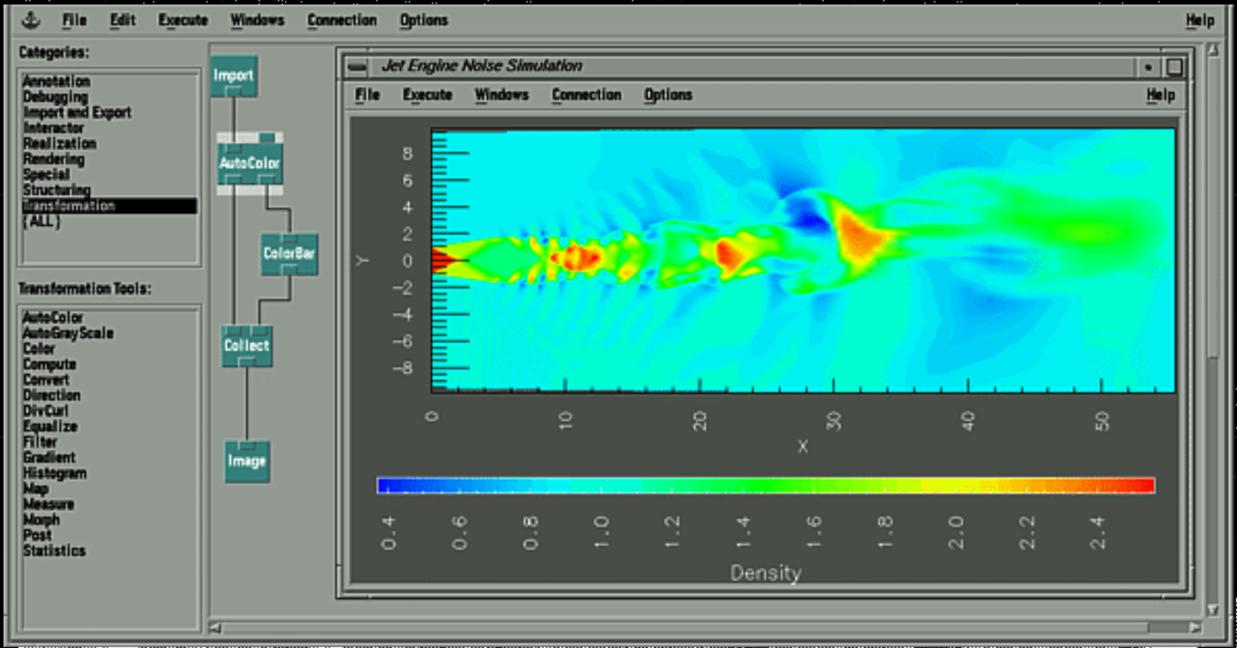}\hfill
\includegraphics[height=\pictureheight]{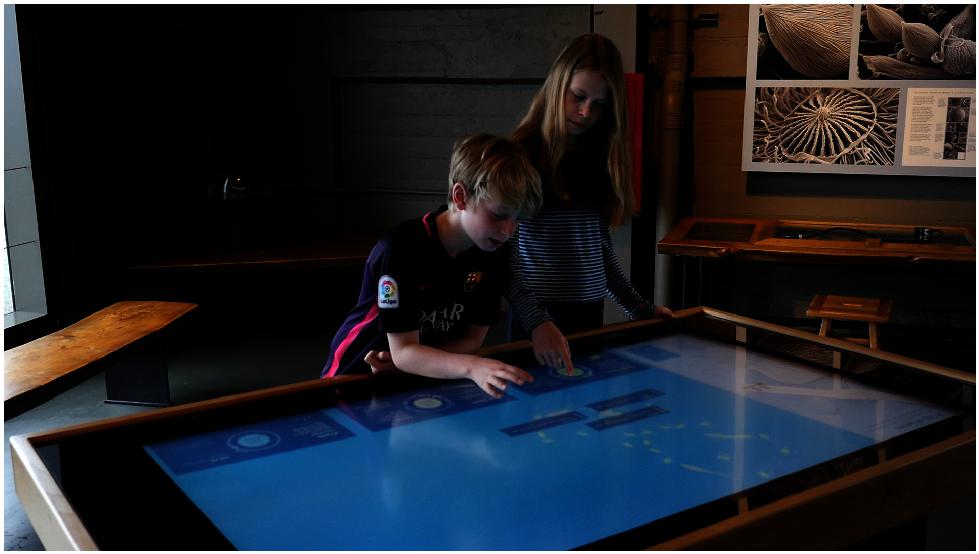}\hfill%
\includegraphics[height=0.17\columnwidth]{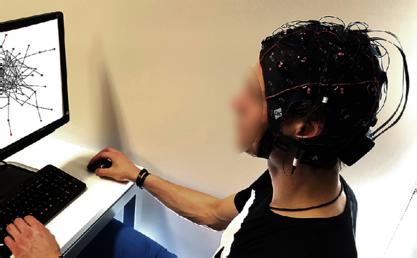}%
\vspace{-2ex}
\label{fig:gui}
\end{figure}



\subsubsection{Schematic Representation and Concept Illustrations}
\label{sec:vc.schematic}
A schematic or concept illustration is an often simplified representation showing the appearance, structure, or logic of a process or concept. Typical examples include 
flowcharts to illustrate an algorithm,  process diagrams, or sketches. 
Schematics and illustrations are common in research papers, not just in visualization papers. 
{We coded  \numschematicOrg schematic or concept illustration representations. This category is most common and among these, $79.3\%$ were 2D.}
Canonical examples of this visualization type taken from
\cite{lee2020data},
\cite{jeong2010interactive},
\cite{isenberg2010exploratory}:


\begin{figure}[h!]
\centering
\vspace{-1ex}
\includegraphics[height=\pictureheight]{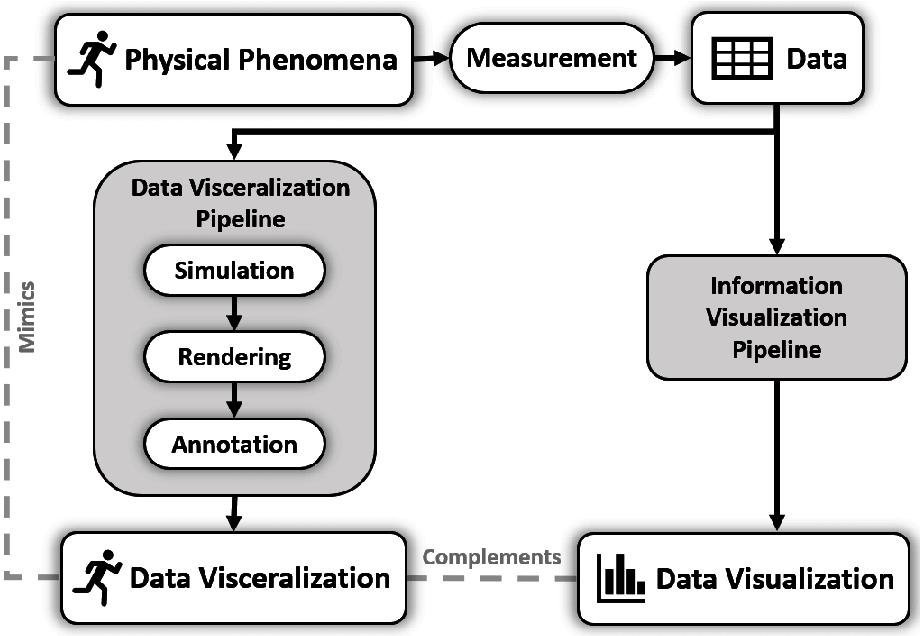}\hfill%
\includegraphics[clip, trim={0 0 0 0},height=\pictureheight]{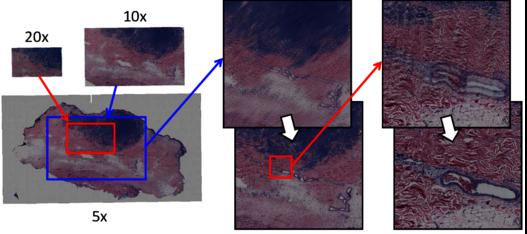}\hfill%
\includegraphics[clip, trim={0 0 1550 0},height=\pictureheight]{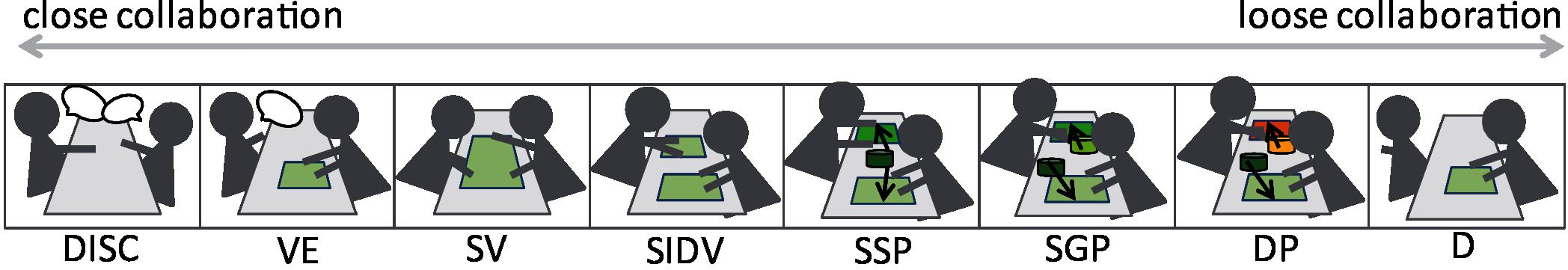}%
\vspace{-2ex}
\label{fig:schematic}
\end{figure}



%



\subsection{Dimensionality: 2D and 3D}
\label{sec:dim}
%
%

Another category we coded concerned the spatial dimensionality of the rendering of  visualizations, GUIs, and schematics.
A flat representation on a 2D plane without perceived depth was labeled as 2D. 
Those images that appeared to be in 3D (or volumetric) space were classified as 3D.
To classify an image as 3D we looked for depth cues such as occlusion, lighting and shading, parallel and perspective projection, rotation, or any other depth cues.
We also required a continuous depth with smooth transitions between depth values.  In other words, we did not generally code images as 3D when the content resided in just one or two 2D planes, \eg, an artificial discrete depth. 
As we shall see in \autoref{sec:ambiguous}, even this apparently simple exercise turned out to be non-trivial for many images due to the presence or absence of a mixture of depth cues.
{Two-dimensional representations became more common than 3D after 2005, perhaps due to the new VAST conference (\autoref{fig:dimDistributionByYear}). 
It is also not surprising that 3D representations were used more broadly for \vissurface (\autoref{fig:typeDistribution}). For the visualization types \vissurface and \visline we observed a decrease of 3D use over time.
}

\section{Challenges On Judging Visualization Types
}
\label{sec:ambiguous}



%

\noindent
While the previous section might seem relatively straightforward to apply, ambiguous cases were much more common than we thought. In this section we focus on discussing our most important challenges during the coding process.

\subsection{Choosing the Right Level \& Members for our Typology}
During out discussions we made several failed attempts at deriving a typology that we could apply to images without knowing details about the data they represented or what construction rules led to them.

\textbf{Bertin's marks and channels as inspiration.}
One of our first attempts was to use Bertin's semiology of graphics and in particular his marks and visual variables for describing visualizations \cite{bertin1983semiology}.  Using this approach resulted in numerous (low-level) codes per image that together did not allow us to meaningfully describe what we saw. 
For example, a scatterplot would be coded as point marks with a position encoding (visual variable). What we wanted to establish were instead higher-level categories that would include both the graphical primitives used and the coordinate system. Still, Bertin's definition of visual variables and marks provided inspiration that we see in the naming of several of our visualization types. For example, we developed a more general `` point-based representations'' category that covers scatterplot-like images. 
In this category, the visual mark of a point would be dominant, perceptually.



\textbf{Visualization Techniques as a Typology?} 
Ward once said ``I'll Know it When I See it''~\cite{ward2002taxonomy}. 
We rely on our observations to help us derive knowledge about data even when we do not know what the data is in detail. Yet, our second failed attempt at arriving a typology began from trying to identify dedicated visualization techniques in the images we saw. We wanted to characterize the ``output space'', the result space of encoding techniques. However, different data encoding techniques could result in similar visuals while the identification of some techniques required knowing the data that was being visualized. For instance, timeline visualizations could only be identified if the axis labels were clearly pointing out temporal data. We later completely abandoned the approach to use visualization techniques for building a typology so that we could focus on what the data representation actually looked like. What we retained from this failed attempt, however, was a focus on the central encoding technique. We intentionally chose not code legends, labels, and embellishments etc.

\textbf{Is ``continuous color-based encoding'' a separate type?}
Several of our visualization types make reference to Bertin's visual channels position and size. For example grid-based encodings rely on specific positions for information layout. Point-, bar-, and area-based representations are distinguished in terms of how size encodes information. Point-based encodings primarily reference a position, bar- encodings a length, and area-based encodings a two-dimensional size. Despite the fact that properties of color such as hue, saturation, and luminance are frequently used to encode data as well, we did not have a dedicated visualization type for color-based encodings. Instead we initially coded most continuous and discrete heat-map type encodings under ``generalized matrix / grid'' with the argument that these encodings are applied on pixel grids. During Phase 6 of our codings, many discussions centered around the question of when a continous color-based encoding should be coded as a grid - especially when it did not look at all like a grid. For example, when continuous colorscale were applied to 3D geometries such as streamlines, the matrix/grid encoding no longer seemed appropriate. Hence, after many discussions, we decided to create a distinct category to recognise the use of continuous colormaps.

\subsection{General Coding Ambiguity}
Many of the challenges with our typology of visualization types stemmed from ambiguities in how we should apply the code set we developed. Here, we discuss the most important of these challenges.

\textbf{Surfaces and volume rendering.}
In an early visualization-type code set we had included surface and volume renderings as two separate codes. Several of us, however, found it very challenging to discern the differences between some volume and surface representation. For example, volume rendering may add thickness but it can also depict surfaces. 
We decided that ``recognizing'' an image type at this level of detail (e.g., whether surfaces are produced 
through volume rendering or surface construction) was not reliable and may not even be 
important in terms of describing a figure's visual content as the underlying algorithmic techniques would be transparent to viewers.  
In machine learning, visualization, and computer graphics, it has been argued that high-level concepts directly contribute to reasoning~\cite{chen2019looks, wang2020toward}. High-level concepts remove the specific details of a given technique and focuses on what all instances of that family of visualization type have in common. We thus chose a visualization type that is more general, i.e., surface and volume
techniques combined.

\textbf{Ambiguous Area-based Images.}
We tried to avoid using data types in our classification schema because we wanted the focus to be consistently on images. We originally had a cartographic map category and choose to remove it because it was both focus on a data type and a specific technique. Instead, we decided that the depiction of areas and their relationships was the underlying principle for cartographic maps but also other related techniques such as area charts, stream graphs, etc. However, there were exceptions: route maps, for example, where lines indicate a direct route, were coded separately as 
networks because the routes encoded topological relationships. 
One difficulty we encountered was the distinction between a ``map'' (cartographical map) and a ``terrain'' (wireframe or surface rendering). 
Conceptually, these were very similar. However, using visual appearance as our guide, map images generally appeared to delineate distinct areas while terrains showed smooth surfaces with evaluations. 
Another difficulty related to maps arose when maps were used as a reference structure for data representations layered on top, akin to how gridlines are used on scatterplots. In these cases we had to derive elaborate procedures to be consisten in our treatment of reference structures. We decided only to code the underlying maps if the visualization would seem to change in meaning or message. However, these decisions were very subjective and resulted in several coding inconsistencies.


\begin{figure}[!t]
  \centering
\subfloat[Schematic.]{
\includegraphics[clip, trim={0 0 0 0}, height=0.4\columnwidth]{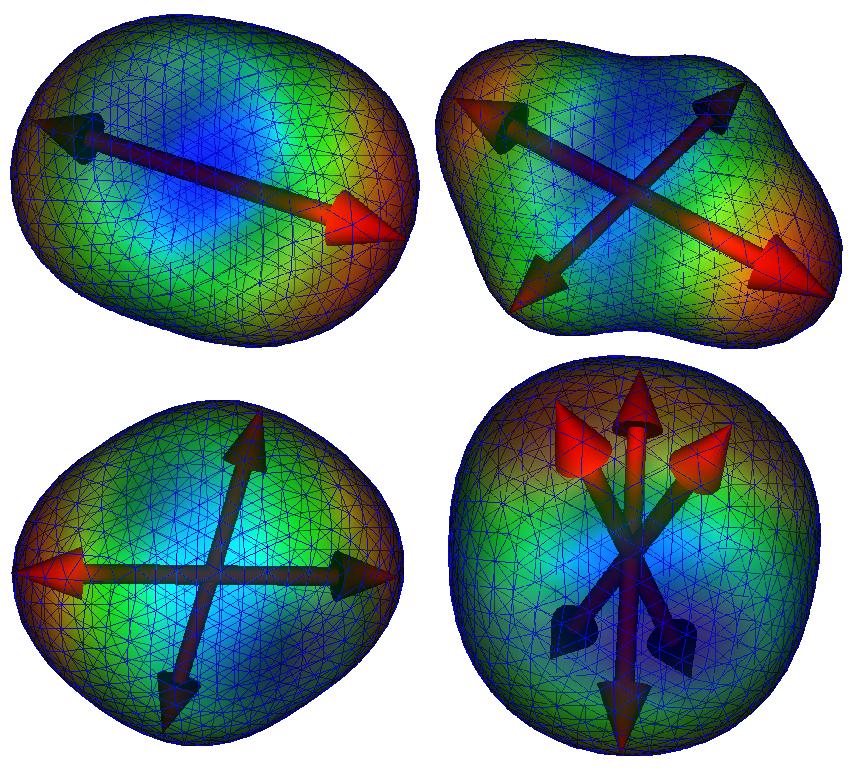}
\label{fig:sa}
}\hfill%
\subfloat[Schematic.]{
\includegraphics[clip, trim={0 0 0 0}, height=0.4\columnwidth]{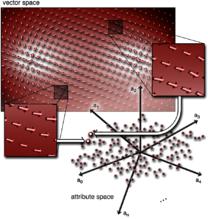}
\label{fig:sb}
}
\caption{Challenging cases of coding \visschematic. (a) Is this a glyph-based or schematic representation? (b) We choose not to code the visualization type inside \visschematic. (Images from (a) Hlawitschka \& Scheuermann~\cite{hlawitschka2005hot} and 
(b) Daniels et al.~\cite{daniels2010interactive}. \textcopyright\ IEEE.)}
\end{figure}

\textbf{Ambiguous Schematic Images.}
A large number of figures were schematic representations or concept illustrations. 
It was often challenging to differentiate between schematics and a demonstration of a visual encoding technique. We often had to abandon our initial goal to ignore what data was encoded to be able to say whether the representation showed a toy dataset. Toy datasets are common in schematic representations, however, the frequent absence of contextual information in images, such as coordinate axis, labels, or scales made the identification of toy datasets difficult. We found that many figures simply did not depict scales which aligns with observations by Cleveland in his review of graphics in other scientific journals~\cite{cleveland1985graphical}. 

We also struggled with the use of annotations in figures as an identification criteria for schematics. For example, ~\autoref{fig:sa} can be coded either as glyph-based in that it shows a mathematical tensor or as schematic to illustrate the authors' design idea or mathematical function. A majority of the team members chose to code \visschematic. Schematic image are often meant to be particularly pedagogical and, thus, include a number of labels and arrows or other annotations. We experimented with a specific coding guidelines in which we considered whether after the removal of annotations we were still seeing an example of a visual encoding type or not---in which case the image would have been a schematic. As such, we agreed that the appearance of labels and annotations to explain an image did not automatically mean the image was a schematic. Our general heuristic for schematics involved establishing if we saw 1) a well-known (or toy) dataset, 2) a pedagogical purpose, and 3) the illustration of a concept.

\textbf{Ambiguous Glyph Cases.}
Glyphs are notoriously difficult to define. Recent definitions emphasize different aspects to delineate a glyph from other encodings. Fuchs et al.\ defined data glyphs as ``data-driven visual entities, which make use of different visual channels to encode multiple attribute dimensions''~\cite{fuchs2013evaluation}.
Borgo et al.~\cite{borgo2013glyph}  followed Ward to define glyphs as ``a visual representation of a piece of data where the attributes of a graphical entity are depicted by one or more attributes of a data record''~\cite{ward2008multivariate}. Munzner's glyph definition is broad and requires a data encoding to be assembled out of multiple marks that encode data \cite{munzner2014visualization}.
For example, each single bar in a stacked bar chart~\wordimg{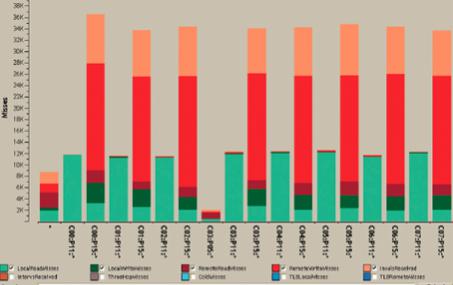} would be  a ``microglyph'' according to Munzner because it is a composite object from multiple length-encoding marks. Throughout our coding we used a mix of the given definitions.

\begin{figure}[!t]
\centering
\subfloat[%
Glyph.%
]{\includegraphics[clip, trim={300 50 1180 0},height=0.4\columnwidth]{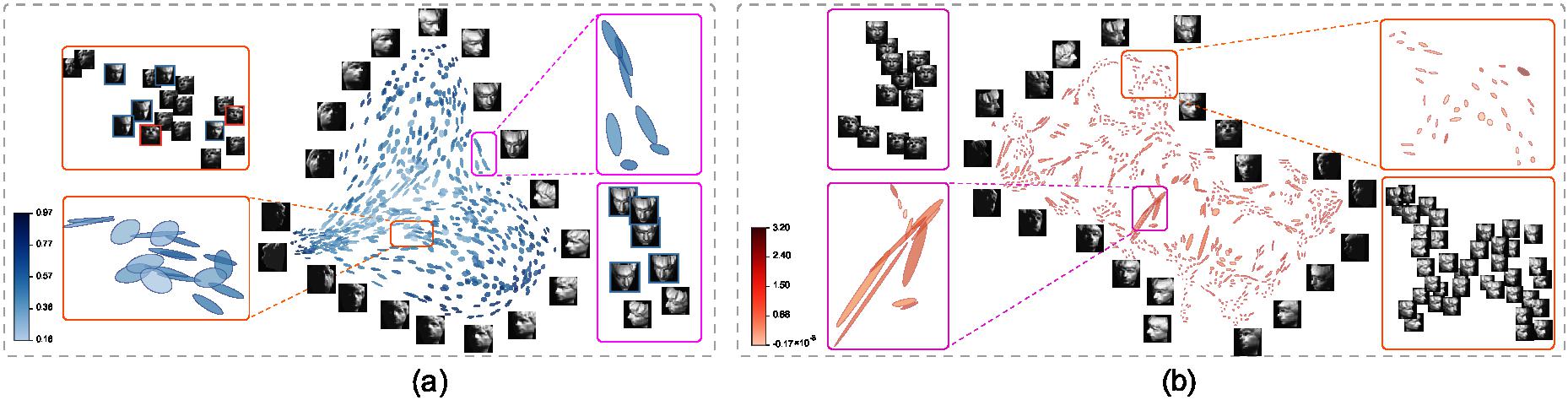}%
\label{fig:glyphOnly}}
\hspace{10mm}%
\subfloat[%
Glyph.%
]{\includegraphics[height=0.4\columnwidth]{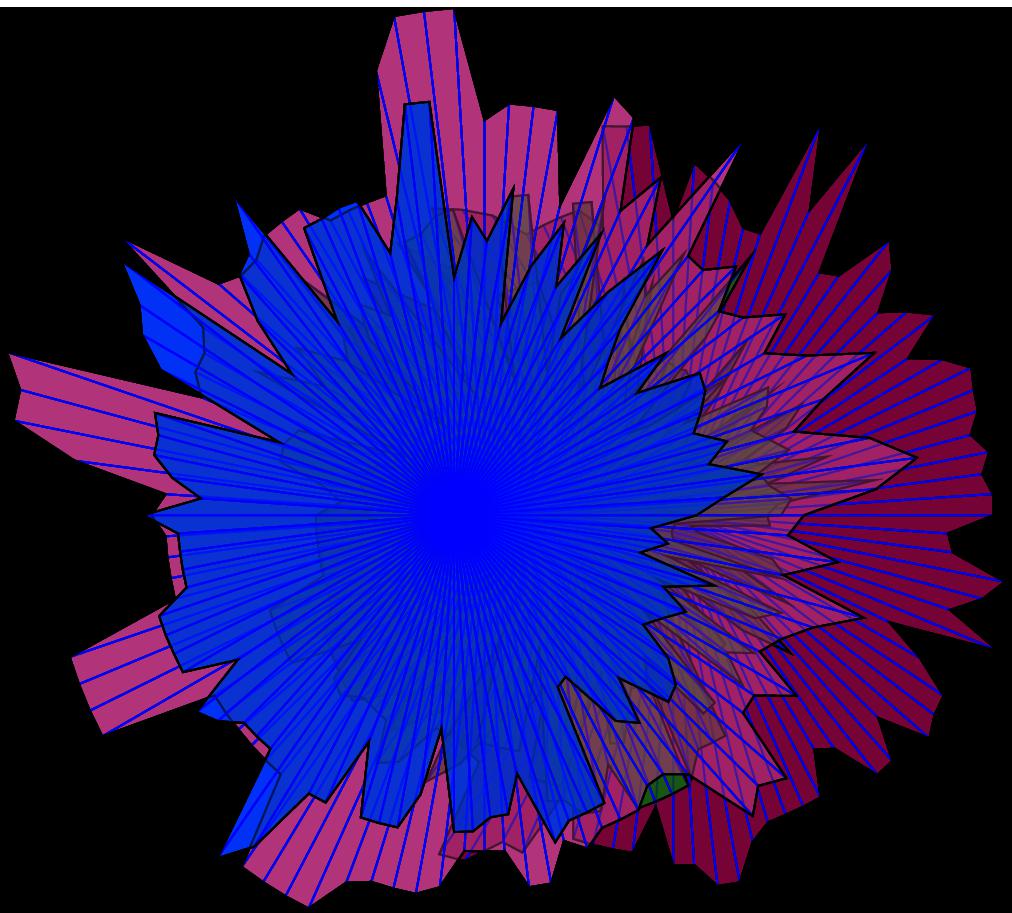}%
\label{fig:areaGlyphglyph}}
\caption{Challenging cases of coding \visglyph. (a) The ellipsoids in the middle could be high-dimensional with two axes and thus orientation and magnitude. 
(b) features starGlyph-like dimensional comparisons and thus is a type of \visglyph.
(Images from (a) Bian et al.~\cite{bian2020implicit} and (b) Fanea et al.~\cite{fanea2005interactive}. \textcopyright\ IEEE.)}
\label{fig:glyphHard}
\end{figure}

Coding \visglyph was challenging as glyphs were often associated with a placement strategy. 
For example, a focus on identifying a specific position is a property they share with point-based techniques. 
However, most commonly glyphs have been defined as representations of multi-variate data which by itself would not help to distinguish glyphs from other visualization types in our typology. 
A challenge is deciding when a point or other graphic primitive becomes a glyph. There is no standard definition to decide after how many data dimensions a single mark becomes a glyph and when a glyph becomes a chart. There seems to be, however, a general consensus that a glyph requires to reach a certain level of complexity to be categorized this way. However, reaching consensus on the level of complexity is challenging. We chose to label an image as \visglyph if there were multiple representation of data points that represented both position and additional data dimensions using color, shape, or other geometric primitives. ~\autoref{fig:glyphHard} illustrates some difficult cases,
when both coders scored the difficulty as ``hard''.

\subsection{Multiple Encoding Ambiguity}

\begin{figure}[!t]
\centering

\subfloat[
Node-link and surface.
]{\includegraphics[clip, trim={680 50 0 0},height=0.3\columnwidth]{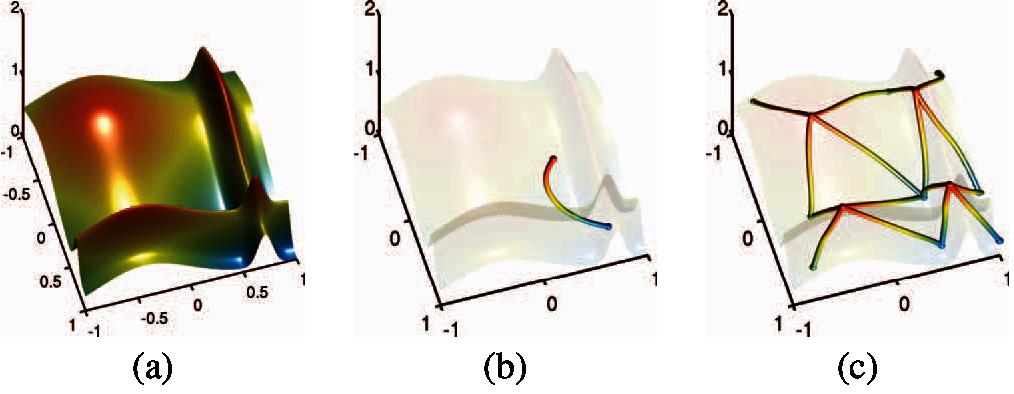}%
\label{fig:nodelinkSurface}}\hfill%
\subfloat[
Line-based, point-based, and surface.
]{\includegraphics[clip, trim={0 0 0 0},height=0.3\columnwidth]{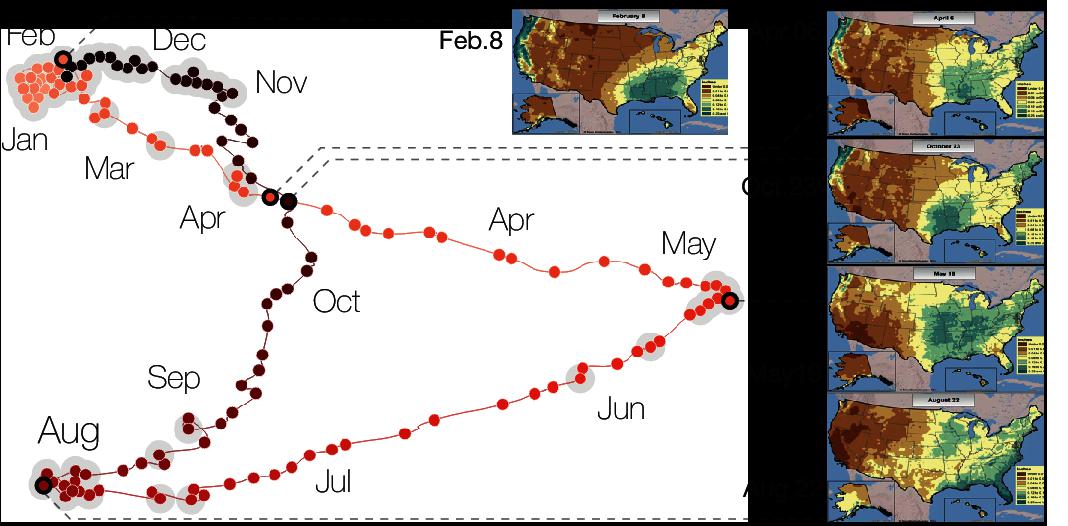}
\label{fig:linepointsurface}}\vspace{-2.5ex}
\caption{Challenging multiple coding ambiguity. Here, the two code signifies different aspects of the data and can be separated to stand alone. As a result, multiple codes apply.
(Images from 
(a) Suits et al.~\cite{suits2000simplification}
and
(b) Bach et al.~\cite{bach2015time}. \textcopyright\ IEEE.)
}
\label{fig:nodelinkMeshes}
\end{figure}

\begin{figure}[!bt]
\centering
\subfloat[
Continuous-color.
]{\includegraphics[height=0.16\textwidth]{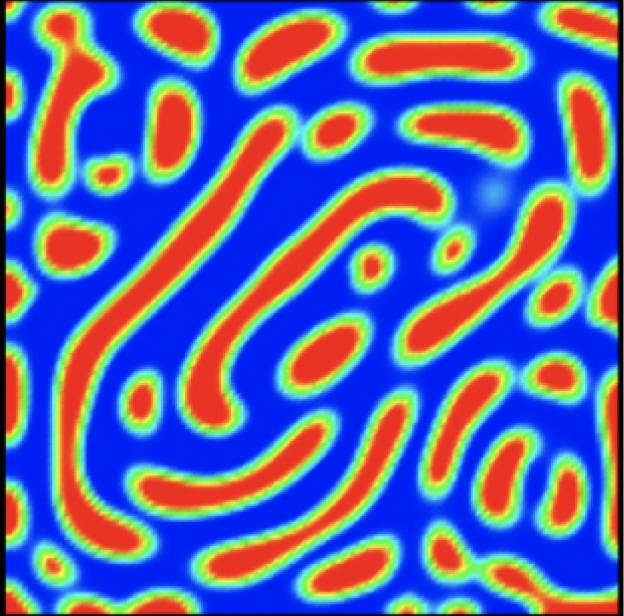}%
\label{fig:c}}\hfill%
\subfloat[
Continuous-color \& glyph.
]{\includegraphics[height=0.16\textwidth]{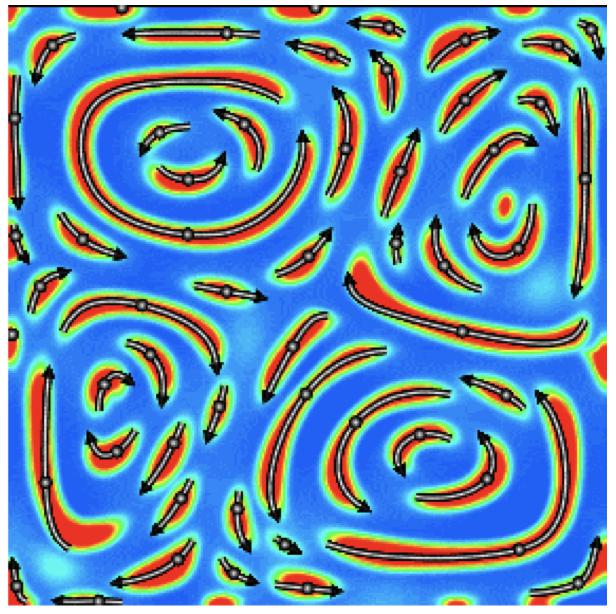}%
\label{fig:gc}}\hfill%
\subfloat[
Pattern-based \& glyph.
]{\includegraphics[height=0.16\textwidth]{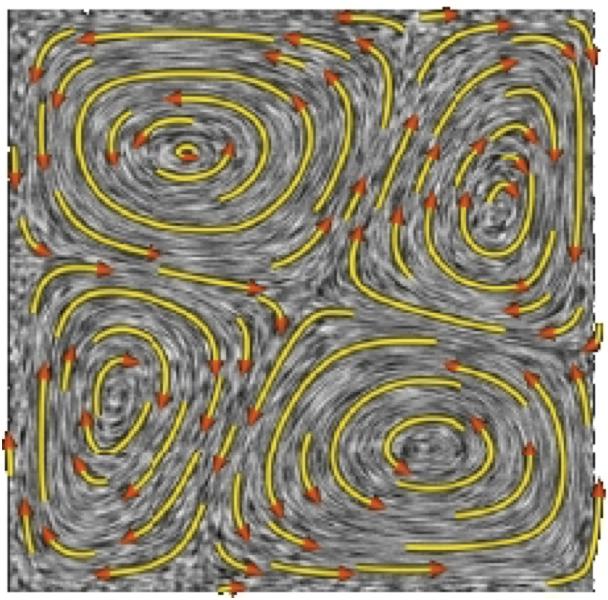}%
\label{fig:fgp}}\vspace{-2.5ex}
\caption{
Challenging multiple coding ambiguity. Here, we did not code \vissurface and only chose the primary code, which differentiates these visualization techniques (Images from 
(a) Weiskopf et al.~\cite{weiskopf2005overview} and 
(b,c) Garcke et al.~\cite{garcke2000continuous}. \textcopyright\ IEEE.)
}
\label{fig:flowcases}
\end{figure}

Many images showed multiple visual encodings which is one of the primary reasons for coding inconsistencies we encountered. We agreed to code multiple visualization types, if these types were distinctive and could be perceived clearly. For example, if there were multiple visual designs layered or nested, that could be distinguished from one another, we tagged more than one encoding, e.g., node-link + area for \autoref{fig:nodelinkSurface}.  We also decided when not to check multiple encodings. For example, the most frequent representation of a confidence interval includes a bar whose length represents the interval and a dot to represent the average. In this case we chose to code the bar as a primary encoding to which the average was considered and annotation. However, these decisions were difficult, often inconsistent, and error-prone (for example, the dot for the average is not merely an annotation when confidence intervals are not symmetric). 



\textbf{Influence of Coder Expertise}
Though we chose to classify images without precise knowledge of the underlying data nor the data type 
(we purposefully ignored the figure captions),
coder expertise often still played a major role in resolving ambiguities. For some images, we were familiar with the original papers, but for most cases of ambiguity, experience may have helped tremendously in understanding the intention behind images; especially for schematics. 
In a sense, our decision of coding an image depended not only on the perceived structure but also the intended function of the image. 
For example, scarf plots~\cite{somarakis2020visual} 
were categorized as matrices rather than point-based encodings because we considered space for lines to be samples along a grid. 

\section{Discussion}

This section presents our reflection on our two-year coding process, limitations, and future work.

\subsection{From Specific Techniques to A General Typology
}

Essentially, our coding experiment evaluated if visualizations (in academic publications) can be easily understood and categorized by experienced researchers. 
While we started with the intuition of finding categories based on data, tasks, and low-level encoding principles (characterization of the ``input space''), we ended up scraping this and came up with a typology of the result or output space of images. 
Some of the original categories survived (point-based, line-based, and generalized area; surface-based and text-based; as well as node-link and glyph) and further new ones have emerged (bar representations, matrix/grid, continuous patterns, continuous color). 
Our typology also combines constructive features (low-level design space construction elements) with functional characteristics from the output space, especially for images that contained schematics and GUIs.

The design space elements are perhaps close to psychophysics---evaluations are well posed and hypotheses can be tested with an (often-lab-controlled) experiment. Resulting images, 
on the other hand, have much more to do with the viewers' knowledge and context, and has its footing in vision science and even machine learning. 
What-we-see could heavily influence how we act to choose to see next, to provide another angle to understand visualization effectiveness in the future.

Furthermore, implicit in teaching and learning tasks such as ``show me the node-link diagrams'' are much deeper issues involving the notion of what is meant by ``node-link diagrams''. 
This meaning would vary with spatial (e.g, topological connectivity) and non-spatial data (e.g., social networks), context of use, and observers. 
We found that student coders involved in the earlier stage of this project had dramatically different understandings of author keywords (often with misconceptions). 
Also, our initial exercise (which lasted for more than a year) of 
measuring \textit{visual design terms of authors or of low-level features} 
has largely been challenged by low-level details of naming techniques rather than what the visualizations show us. 

%
%

\subsection{
What We See and Speed of Human Reasoning}

The most significant contribution of our work is the derivation of a small set
of categories that attempt to cover all techniques (completely). 
By merely seeing a figure, we purposely avoid focus on the data but rather focus on  interpretation of visual design alone. 



Our typology might be the first typology easily accessible
to draw clear boundaries between visualization image types.
Intuitively, a bar chart can be encoded as position or height, but users may not relate them to low-level taxonomy. 
Describing a bar chart as position and length may not be as accessible to the general audience as it is for a visual design expert. 
Similarly, it is easier to describe a histogram as a more typical \vislength than a ``distribution plot''.
Murphy calls effects like this `typicality'~\cite{murphy2004big} for any task requiring relating an item to its categorical concepts, using typical terms encourages learning and usefulness for inference tasks. 
These categorical concepts further
explain how we can understand objects we have never seen before.
and extrapolate new categories from a few given examples.
This current work may partially reflect this, since the expert coders could not agree on detailed categories when specific techniques were used.
Expert coders also could not always visually distinguish between techniques.
Some reasoning, such as volume rendering is fuzzier than surface rendering, does not fit all cases and coders
did not feel confident in
categorizing these.
Even volume graphics experts made frequent mistakes except for those images they knew 
a priori. 
Coding accuracy is heavily influenced by the coder's own expertise and experience in a given area. 
Such a disagreement suggests that visualization techniques may not be as easily accessible as we think. 



\subsection{Limitations and Future Work}

Describing visual images at this what-we-see level--matching human intuition has not been studied in-depth previously. 
Tufte mentioned that ``when principles of design replicate principles of thought, the act of arranging information becomes an act of insight''(\cite[p. 9]{tufte1998visual}). 
The difficult question is whether we can ask visualization scholars to imitate this way of reasoning: to interpret an image in connecting a visualization to the designers' intent. The potential next step is to invite the community to use our typology to study / refine categories. 

The evaluation of our typology is done through its methodology 
(initial coding pass, multiple coders resolve conflicts, codes being iterated and discussed over the course of approximately two years). 
Additional validation, e.g. through external researchers that were not part of the coding team, would be a good next step, but is outside the scope of this paper. 
Our framework might potentially be used to compare and analyze what humans and machines see differently from these visualization types.  

\section{Conclusion}
This article reflects on our journey to define a new visualization typology using high-level categories.
The journey began with community-defined visualization keywords. Two failed attempts to use technique keywords and low-level encodings finally results in our typology of visualization types that focuses on describing our community's ``output''. Our visualization types emerged from both structural and functional similarities of the images. 
Indeterminacy of hard cases reflects perceptual uncertainty but similarity in each category---this looks like that---gives insight into the understanding of visualization techniques. 
The typology developed here could provide a potentially powerful framework for studying topics of interests in visualization, such as image retrieval.



\bibliographystyle{abbrv-doi-hyperref-narrow}

\bibliography{ms}

\appendix

\clearpage
\onecolumn
\noindent\begin{minipage}{\textwidth}
\vspace{1cm}
\makeatletter
\centering
\sffamily\LARGE\bfseries
\mytitle\\[1em]
\large Additional material\\[1em]
\makeatother
\end{minipage}
\vspace{1cm}

\noindent While the main document contains 
the main aspects of employed techniques and results, this supplemental
material aims at providing exhaustive and reproducible experimental details.

\section{Input from This Work: Figures} 

The first six years of the seven year's data come from the collection of figure and table data from IEEE VIS publications from 1990--2019. 
We added the 2020 dataset to VIS30K data collection~\cite{Chen2020VIS30Kdata} using their model. In doing this, we reused the authors' approach and their meta-data to remove all tables (unless part of a figure) and their open-source model and tools to 
extract and clean the new figure data of 2020, after
scraping all papers from the IEEE site. 
The first author collected and cleaned the image data manually afterward. The final image set was checked by other VIS30K co-authors.
After collecting these source data, we determined a criterion for visualization characterization that underlies our quantitative analysis of figure content analysis. We coded, refined, recoded, validated, and analyzed the figure content (see
\autoref{tab:12schema} for the final types).

\section{On the Analysis of Judging Visualization Types: Representation Transitions and Heuristics (Additional Results)}
\label{sm:process}

\textbf{The top-21 keywords by authors and the four functional purposes of the images.}
The top-21 keywords are  both \textit{frequently occurring} and \textit{distinguishable} from one another.
This initial process resulted in the following list. 
\textit{bar chart}, 
\textit{cartographic map},
\textit{circular node-link tree}, 
\textit{flow chart}, 
\textit{flowline}, 
\textit{glyphs}, 
\textit{heatmap}, 
\textit{isosurface rendering}, 
\textit{line chart}, 
\textit{matrix}, 
\textit{node-link diagram}, 
\textit{parallel coordinate plot}, 
\textit{pie chart}, 
\textit{point cloud}, 
\textit{scatter plot}, 
\textit{tag cloud}, 
\textit{timeline}, 
\textit{treemap}, 
\textit{volume rendering}, 
\textit{Voronoi diagram}, and 
\textit{wireframe rendering}. An ``other'' tag is added to collect any other techniques that are not on the list.

Our initial purpose classification featured four categories: 
\textit{F1. rendering illustrating a visualization technique}, 
\textit{F2. result presentation from a quantitative evaluation}, and 
\textit{F3. screenshot of system or graphical user interface (GUI)}. 
\textit{F4. photo of a real-world physical scene}, 
These functions are chosen to mainly identify quantitative charts 
(F1) which have a long history of basic research development, 
(F1) algorithm results largely from scientific visualization communities 
(F3) techniques from VAST interface papers, and
(F4) photo used to show real-world imagery.  
Images not in these categories were added to the ``other'' category.

\subsection{Code Changes Between Phases}
The birth and death of these techniques and the transition to typology are shown in other supplemental materials.
We have shared Instructions used by coders in other supplemental materials. 

\subsection{Additional Coding Choices: Concept evolution, rationale, self-correction, and agree to disagree: Early Coding 2006}
\label{sec:heuristics}

\textbf{Not repeatable when characterizing by
Authors' keywords.}
There is also a scalability issue, in that when a new technique is created, a new keyword would be added. It is not repeatable or even findable of these new solutions. In addition, 
focusing on specific cases could only tell us which specific instances of 
visualization techniques were used, but it did not identify cases to which the parts papers used were similar (e.g., both surburst and traditional bar chart ask people to reason with bar height.) Taxonomy is meant to be high-level describing a category of work but this does not exist. 
A good classification must be complete (to cover all designs possible) and relatively clean in that many designs fall into
one category~\cite{gleicher2011visual}. 
Organizing techniques into a hierarchical form to fit new techniques can facilitate learning and searching for popular techniques as well comparing the number of distinct techniques and related uses.

{\textbf{We merge some categories if they contain common visual channels.}} 
For example, flowlines, parallel coordinate plot, line chart all contain line drawings. 
{\codingdecisions{We merged them into the \visline category.}}
The flow chart category was prevalent and represent the largest number of figures in the other category from our 2006 coding, was thus added and later further expanded to \visschematic. The original flow chart has only captured limited number of wiring diagrams. Many charts and diagrams in visualization papers are visually rich containing techniques to illustrate concepts. 
\codingdecisions{We expanded flow chart to ``General Schematic Representation, schematic images, schematic concept illustration''.}

\textbf{We ignored the drawing media used to visualize the data.}
Equally challenging is whether or not views are hand-drawn or computer-generated. For example, schematic ones can be drawn by algorithms~\cite{kim2010evaluating} and there used to be a research field known as ``illustrative visualization'' which developed algorithms that would render images in a style that appeared to be hand-drawn.
The illustrations in the classical book of Bertin's semiotics were also drawn by hand. 
Hence, we chose to ignore the media in the subsequent coding phase.
Instead, we \codingrules{emphasize the elements in the figure rather than the drawing media}. For example, a photograph of an environment (say a VR installation) would be coded as ``3D''. 

\textbf{We managed annotation, legend, and context.}
One of the challenges was the treatment of context information, such as annotation marks, or color legends. Here, we simply focused on the primary visual encoding and agreed to not code such context information separately. Color legend is in ``other'' category. 
Context unless is relevant to the data is not coded. 
Some contextual data, such as geometry models or boundary conditions are important to understand the visualization techniques and were thus coded. 

\textbf{We avoided including data types in the type names.} For example, scalar, vector, and tensor field visualization techniques can be defined using our typology without specifically mentioning flow fields or tensors. 
Thus, we removed this category. 
Text-based encodings is the only exemption to our premise of not coding data-type. 
However, it is such an esoteric type, that we felt it was justified in that case.
We also believe that, in general, users of our typology will want to see this as a distinct category.

\begin{figure}[!t]
\includegraphics[height=.23\columnwidth]{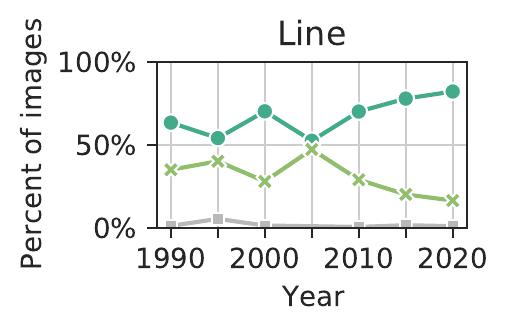}
\includegraphics[height=0.23\columnwidth]{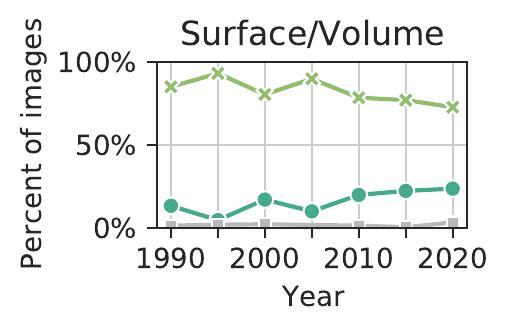}%
\vspace{-10pt}
\caption{Temporal overview of the proportions of 2D and 3D images for \vissurface and \visline.}

\label{fig:surfaceline3D}
\end{figure}

\textbf{Additional Results on 2D and 3D Uses for \vissurface and \visline} are in \autoref{fig:surfaceline3D}.
We see a decline of 3D use over time.








\begin{figure}[!t]
\centering
\subfloat[]{\includegraphics[height=0.2\columnwidth]{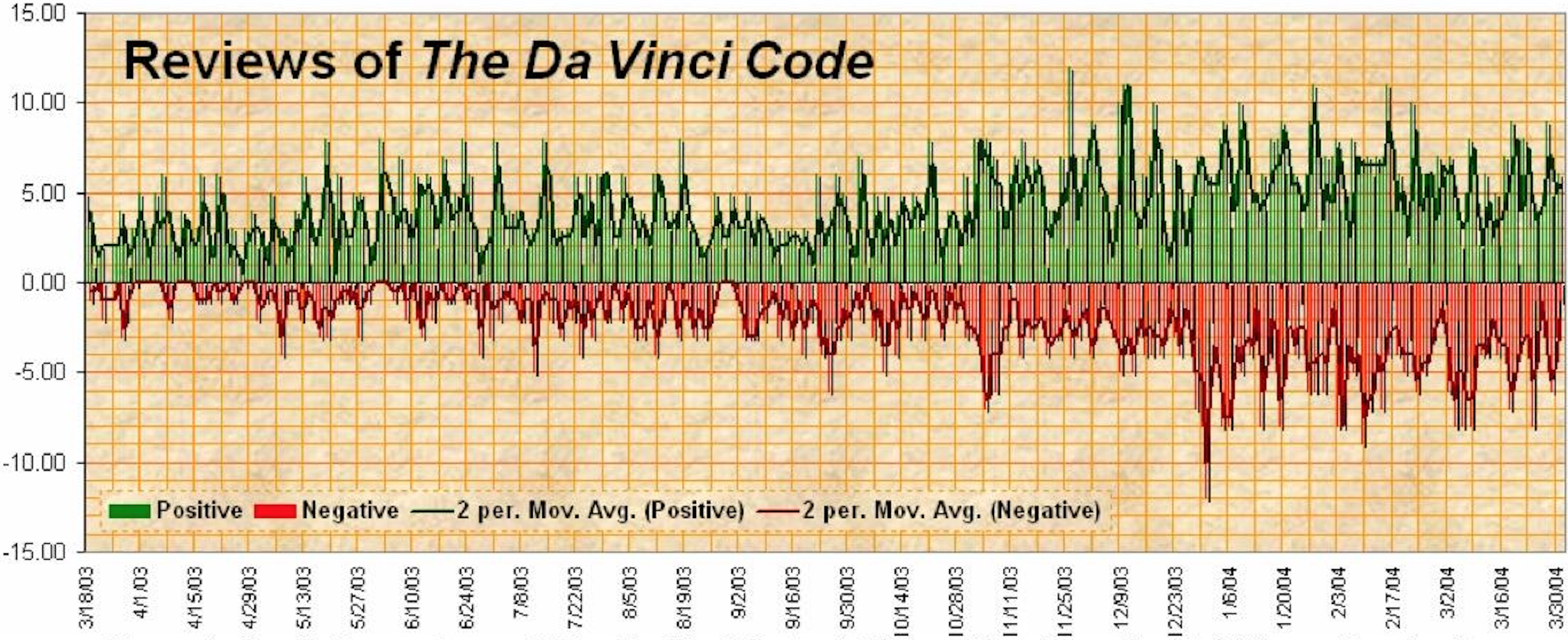}%
}
\hfil
\subfloat[]{\includegraphics[clip, trim={0 30 0 0},height=0.2\columnwidth]{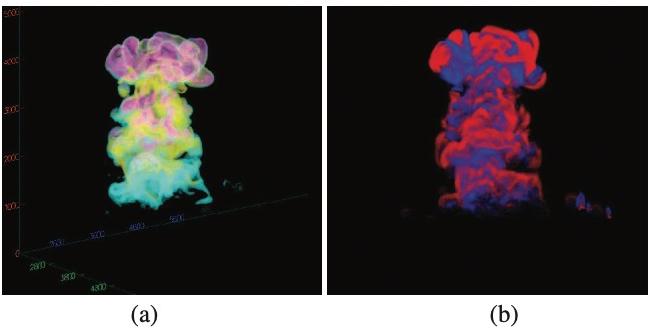}%
}
\caption{\textbf{Rationale to remove timeline from our typology schema.} Is there a timeline in the figures? (a) yes because of the dates on the x-axis and (b) yes but not so obvious and the judgement counts on the coders' knowledge. (Image courtesy of (a) Chen et al.~\cite{chen2006visual} and (b) Song et al.~\cite{song2006atmospheric}.) \textcopyright\ IEEE.}
\label{fig:flowcases}
\end{figure}

\section{Web Interfaces for Annotating Figures}
\label{sm:web-interfaces}



\begin{figure*}[!h]
	\subfloat[]{\includegraphics[clip, trim={50 0 0 0}, height=0.3\textwidth]{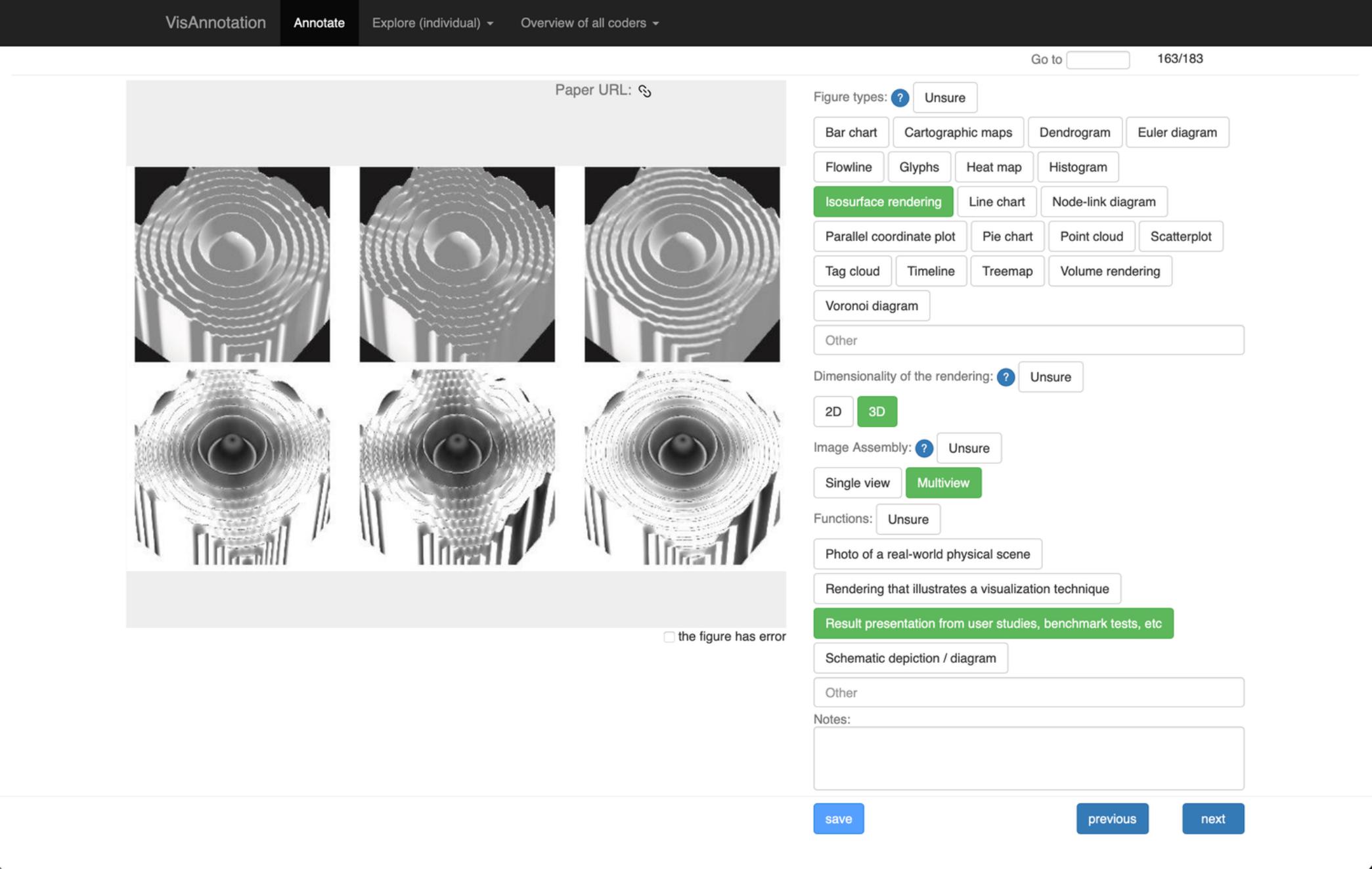}
	\label{fig:web-interface-phase1}}
	\subfloat[]{\includegraphics[clip, trim={50 0 0 0}, height=0.3\textwidth]{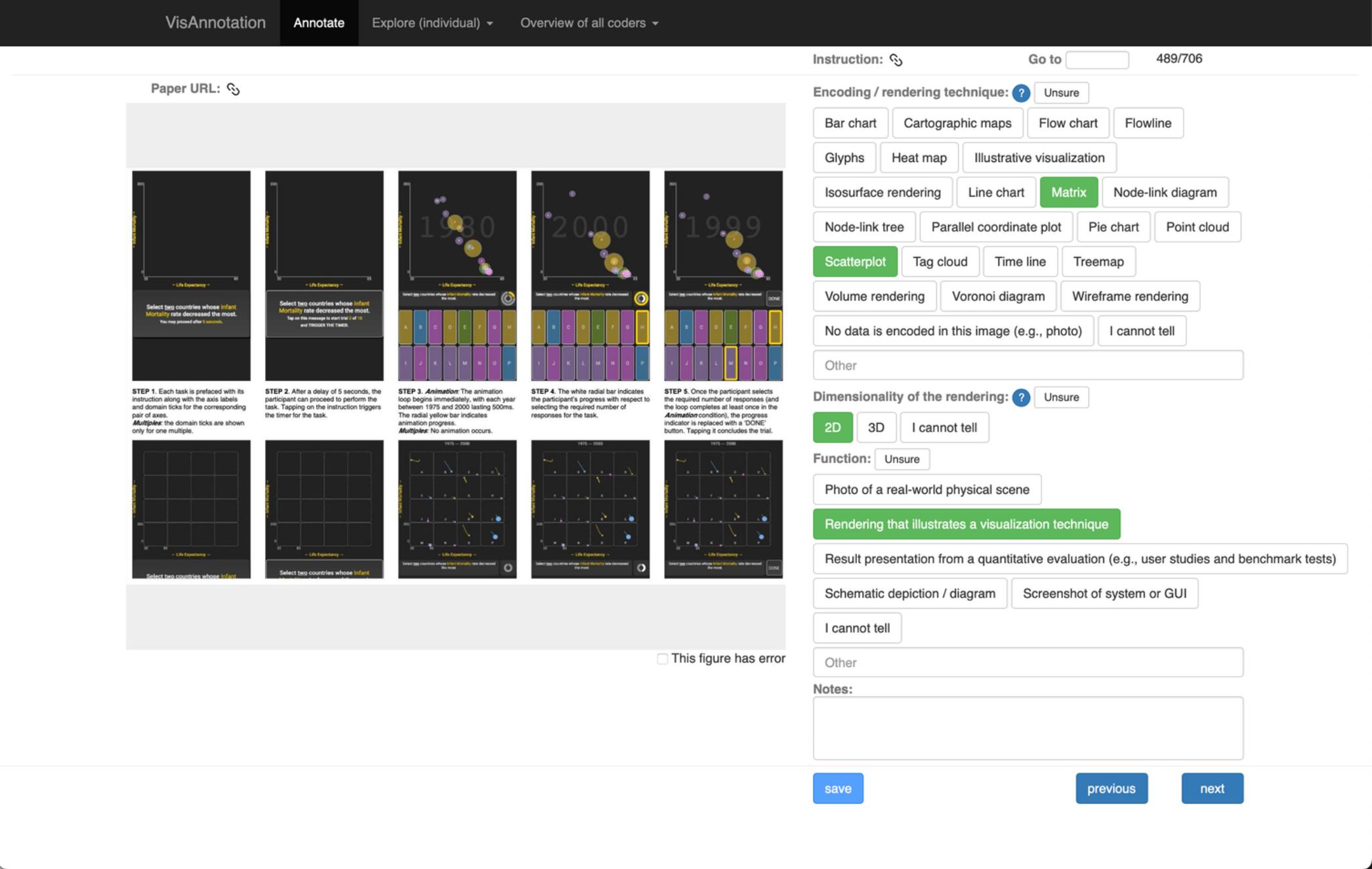}
	\label{fig:web-interface-phase2}}
	
	\subfloat[]{\includegraphics[clip, trim={50 0 0 0}, height=0.3\textwidth]{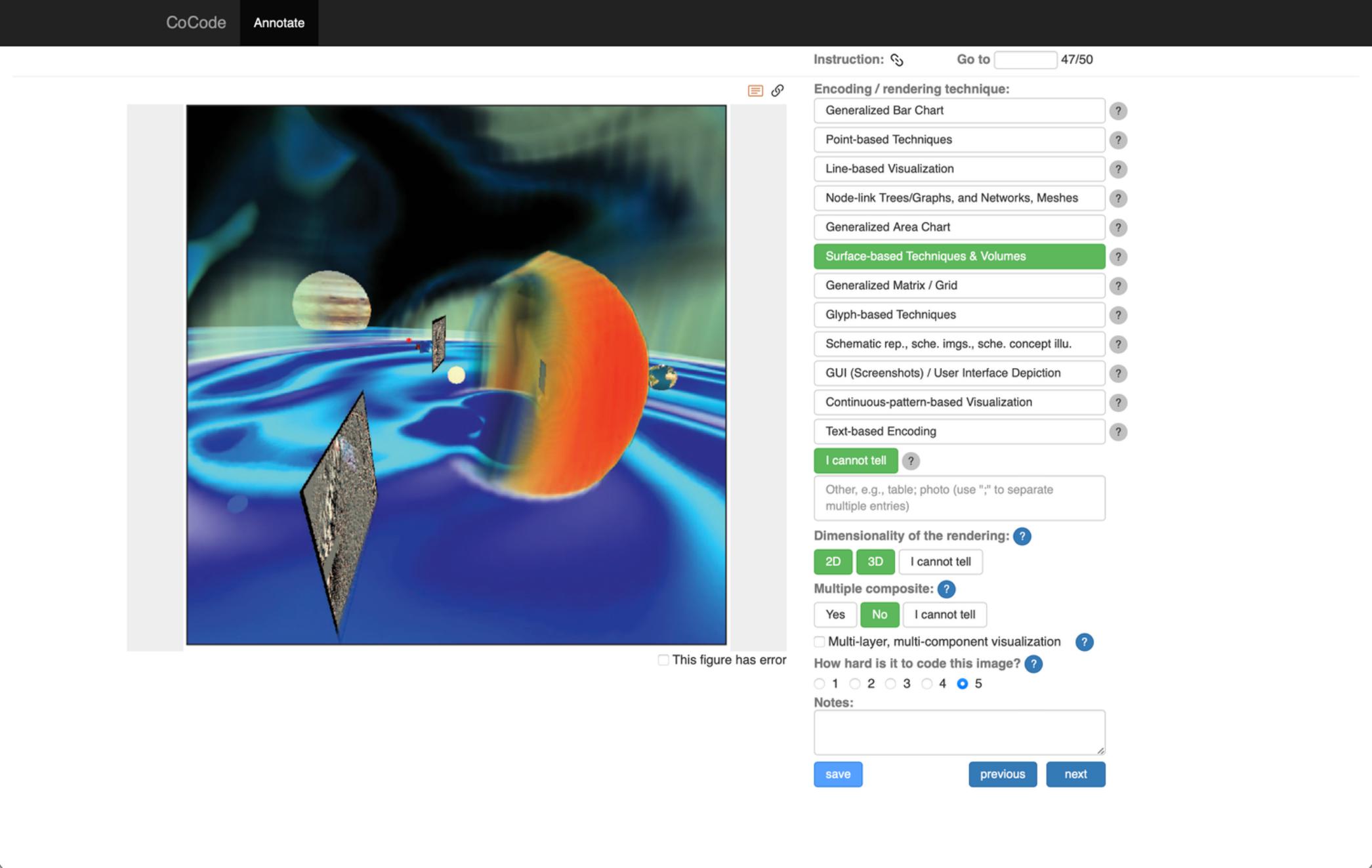}
	\label{fig:web-interface-phase3}}
	\subfloat[]{\includegraphics[clip, trim={50 0 0 0},height=0.3\textwidth]{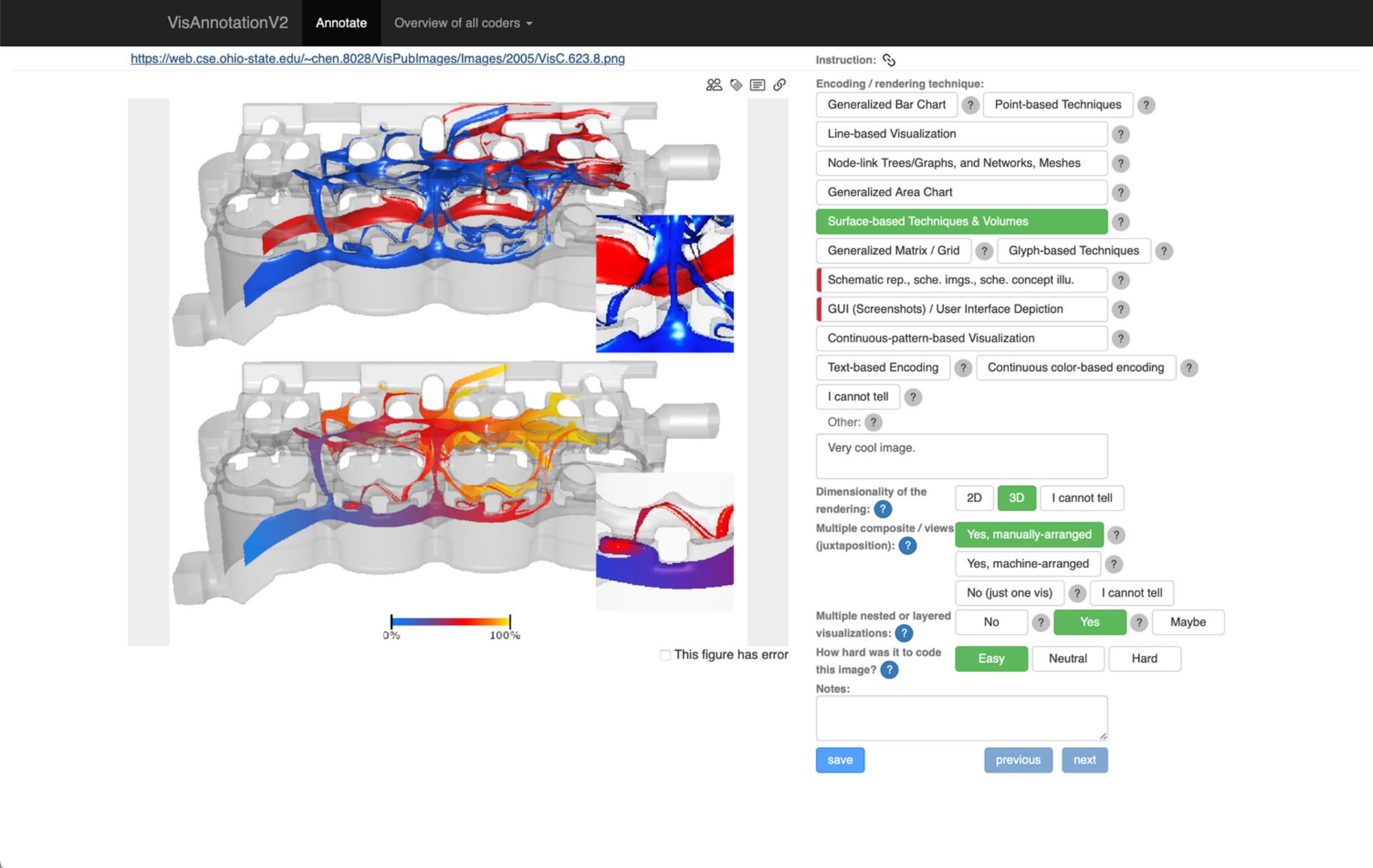}
	\label{fig:web-interface-phase4-5}}
	\caption{\tochange{Screenshot of Phase-1-2-3 and in Phases 4--5 Web interfaces. Phase 1 focused on techniques derived from authors' keywords:  The coders use the interfaces to code each image. The coding labels and categories were updated reflecting the results of weekly discussions. These label interfaces show that code revolution over time: from
	\textbf{Keyword-based} to 
	\textbf{Type-based} updates.
	}}
	\label{fig:web-interfaces}
\end{figure*}

\begin{figure*}[tb]
	\centering
	\includegraphics[height=0.4\textwidth]{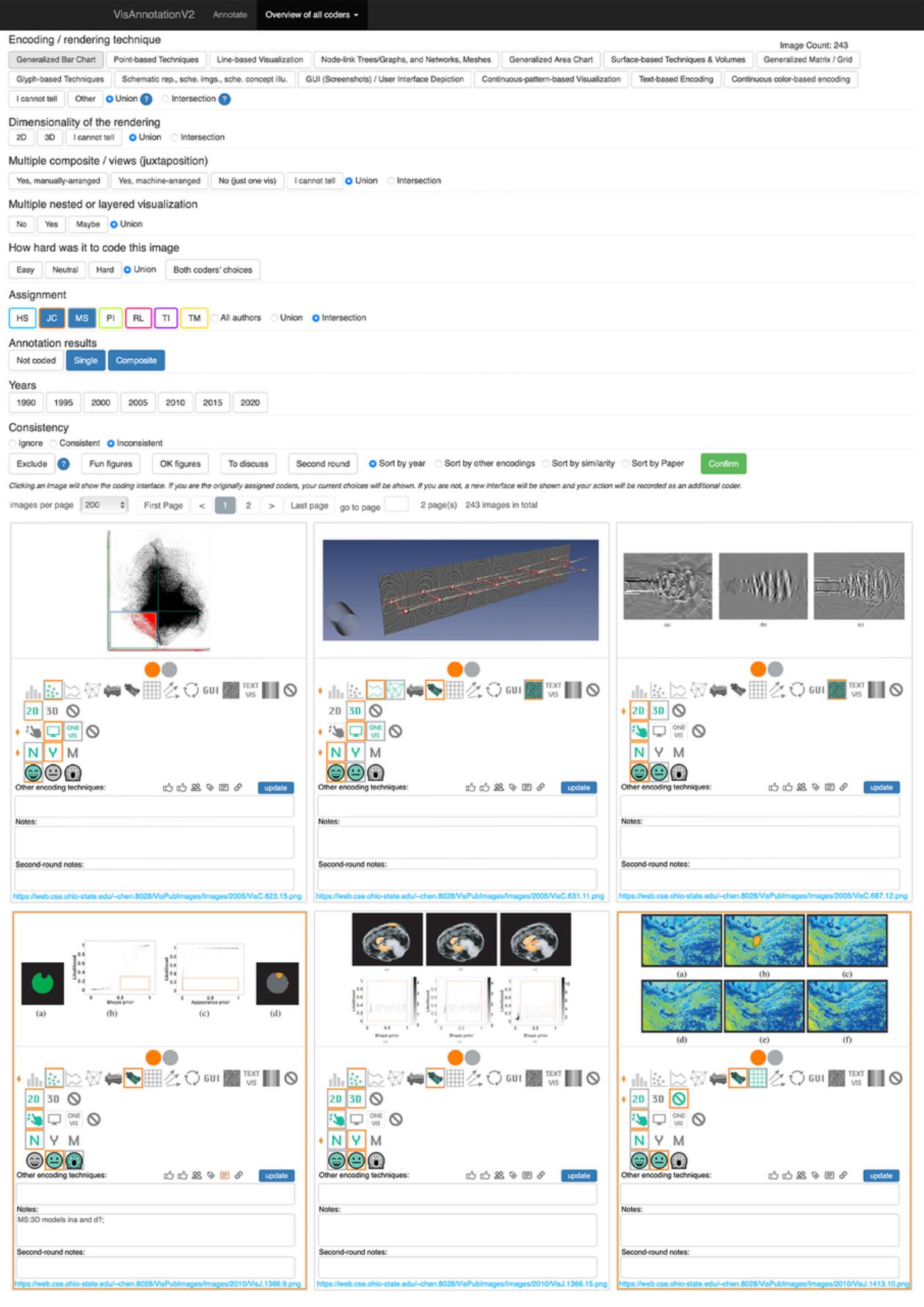}
	\caption{\tochange{Screenshots of Web Interfaces for the Coding in Phase 6 and for Resolving Coders' Inconsistency through Pair-wise Comparison. Two coders results are shown together. For complex images, coders can also look up the images in the same paper coded by other coders through paper-based or image similarity-based search.} 
	}
	\label{fig:web-interface-consistency}
\end{figure*}

We annotated all images  and discussed and compared our codings via our own web-based interfaces to support our collaborative work (\autoref{fig:web-interfaces} (a)-(d)).
Our web-based tool automatically loads images and authorize the uses.
The uses can tag the given image according to the terms, either keyword-based or type-based typology.
On the back-end of our coding tool, we recorded every button click from each participant during the coding process for post-hoc analyses.
For resolving code-consistency, we also implemented a comparative interface (\autoref{fig:web-interface-consistency}) for us to resolve all coding conflicts. Again, all coders' button clicks were recorded.


\section{A Visualization Characterization Tool}


\begin{figure}[!t]
\centering
\includegraphics[width=\columnwidth]{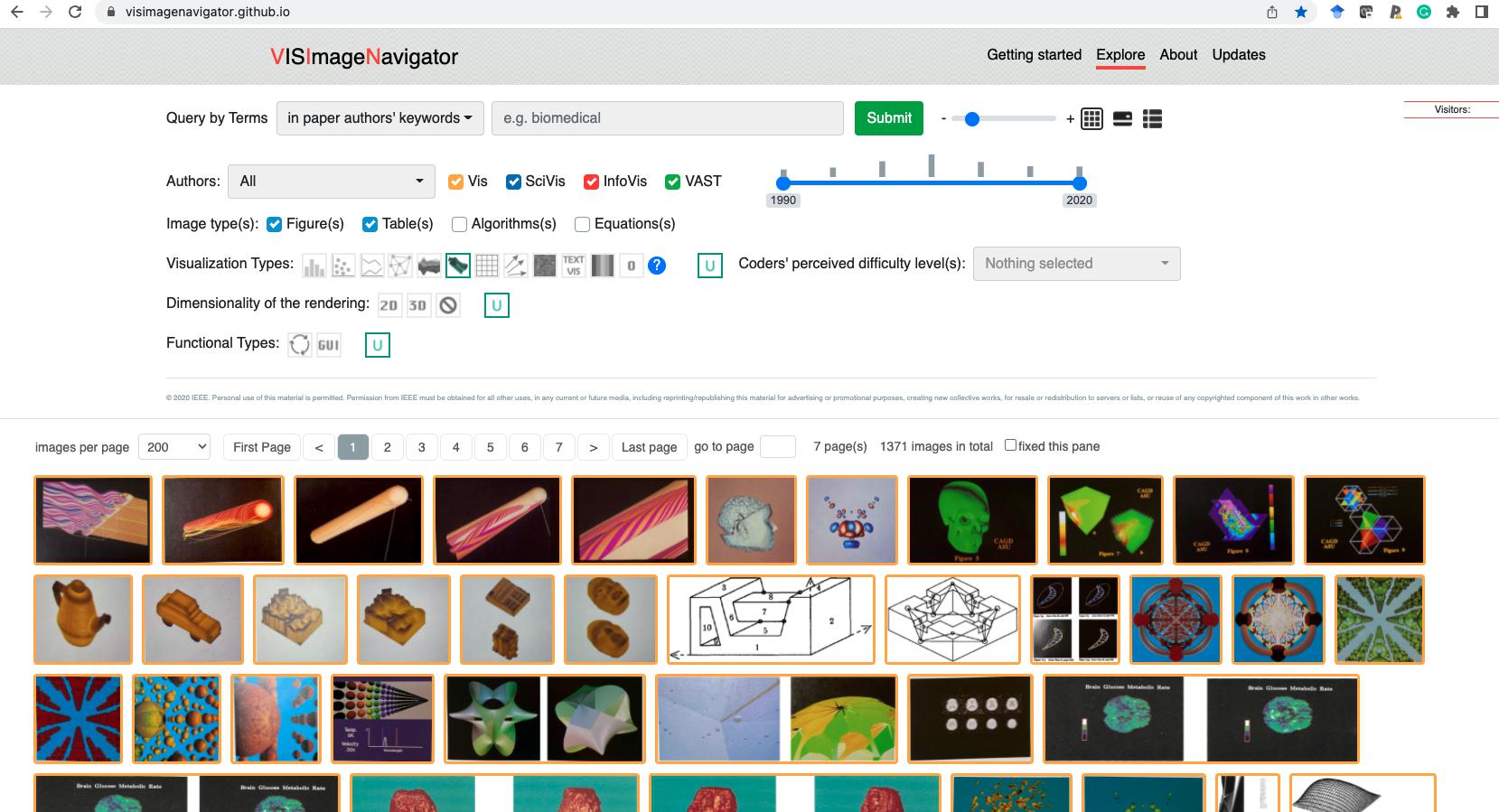}
\caption{Our exploratory tool is integrated with the VisImageNavigator (https://visimagenavigator.github.io/). Viewers can explore our codes of \totaltypelabels labels by visualization types, functional types, dimensionality, and difficulty levels.}
\label{fig:visTypeNavigators}
\end{figure}

We built an exploration tool as an extension of VisImageNavigator (https://visimagenavigator.github.io/) (\autoref{fig:visTypeNavigators})
for viewers to lookup, learn, and
re-evaluate our 13 encoding categories and their dimensionality.
We have augmented the exploration experience with 
\tochange{image similarity}, year, authors, and terms from paper abstract 
and keywords. 

\textbf{Explore Morphological Structures and Form to Design New Techniques.}
A figure may contain many pieces of information. To understand and design one's own, attention must be directed to the 
key uses of techniques and visual details. Within a category 
of our high-level encoding, we may explore morphological variations in structure and form we purposely avoided in this work. 

\textbf{Advance Education and Improve Visual literacy.}
Since our categories are high-level, they should be easy to learn and memorize, thus a very exciting use is to dedicate it to educational uses. We can also use it to explore techniques in specific research areas. Some images and techniques, e.g., photos, might be more meaningful.

\end{document}